\documentclass[twocolumn]{aastex631}
\usepackage{amsmath}
\usepackage{color}

\def\ltsima{$\; \buildrel < \over \sim \;$}
\def\simlt{\lower.5ex\hbox{\ltsima}}
\def\gtsima{$\; \buildrel > \over \sim \;$}
\def\simgt{\lower.5ex\hbox{\gtsima}}
\def\g{{\rm\,g}}

\def \r {r$^{1/4}$ }


\def\s{\ifmmode \widetilde \else \~\fi}
\def\={\overline}

\def\spose#1{\hbox to 0pt{#1\hss}}

\def\lta{\mathrel{\spose{\lower 3pt\hbox{$\mathchar"218$}}
     \raise 2.0pt\hbox{$\mathchar"13C$}}}
\def\gta{\mathrel{\spose{\lower 3pt\hbox{$\mathchar"218$}}
     \raise 2.0pt\hbox{$\mathchar"13E$}}}
\def\Dt{\spose{\raise 1.5ex\hbox{\hskip3pt$\mathchar"201$}}}    
\def\dt{\spose{\raise 1.0ex\hbox{\hskip2pt$\mathchar"201$}}}    

\def\r{${\rm r^{1/4}}$}

\def\ltsima{$\; \buildrel < \over \sim \;$}
\def\gtsima{$\; \buildrel > \over \sim \;$}
\def\lsim{\lower.5ex\hbox{\ltsima}}
\def\gsim{\lower.5ex\hbox{\gtsima}}
\def\lapp{\ifmmode\stackrel{<}{_{\sim}}\else$\stackrel{<}{_{\sim}}$\fi}
\def\gapp{\ifmmode\stackrel{>}{_{\sim}}\else$\stackrel{<}{_{\sim}}$\fi}

\definecolor{applegreen}{rgb}{0.55, 0.71, 0.0}

\newcommand{\vsig}{\mbox{$v/\sigma$}}
\newcommand{\simba}{\mbox{{\sc Simba}}}
\newcommand{\gizmo}{\mbox{{\sc Gizmo}}}
\newcommand{\gad}{\mbox{{\sc \small Gadget-3}}}

\accepted{ApJL}

\shorttitle{The universality of the HI-to-Halo mass ratio for isolated disk galaxies}
\shortauthors{Korsaga et al.}
\usepackage{amsmath}

\begin{document}

\title{Disk galaxies are self-similar: the universality of the HI-to-Halo mass ratio for isolated disks}

\correspondingauthor{Marie Korsaga}
\email{marie.korsaga@astro.unistra.fr}



\author[0000-0002-5882-610X]{Marie Korsaga}
\affiliation{Universit\'e de Strasbourg, CNRS, Observatoire astronomique de Strasbourg, UMR 7550, F-67000 Strasbourg, France}
\affiliation{Laboratoire de Physique et de Chimie de l'Environnement, Universit\'e Joseph Ki-Zerbo, 03 BP 7021, Ouaga 03, Burkina Faso}
\affiliation{Instituto de Astrof\' isica de Canarias, Calle V\' ia L\'actea, s/n, E-38205, La Laguna, Tenerife, Spain}

\author[0000-0003-3180-9825]{Benoit Famaey}
\affiliation{Universit\'e de Strasbourg, CNRS, Observatoire astronomique de Strasbourg, UMR 7550, F-67000 Strasbourg, France}
\author[0000-0002-5245-7796]{Jonathan Freundlich}
\affiliation{Universit\'e de Strasbourg, CNRS, Observatoire astronomique de Strasbourg, UMR 7550, F-67000 Strasbourg, France}

\author{Lorenzo Posti}
\affiliation{Universit\'e de Strasbourg, CNRS, Observatoire astronomique de Strasbourg, UMR 7550, F-67000 Strasbourg, France} 

\author[0000-0002-3292-9709]{Rodrigo Ibata}
\affiliation{Universit\'e de Strasbourg, CNRS, Observatoire astronomique de Strasbourg, UMR 7550, F-67000 Strasbourg, France}

\author{Christian Boily}
\affiliation{Universit\'e de Strasbourg, CNRS, Observatoire astronomique de Strasbourg, UMR 7550, F-67000 Strasbourg, France}

\author{Katarina Kraljic}
\affiliation{Universit\'e de Strasbourg, CNRS, Observatoire astronomique de Strasbourg, UMR 7550, F-67000 Strasbourg, France}

\author{D. Esparza-Arredondo}
\affiliation{Instituto de Astrof\' isica de Canarias, Calle V\' ia L\'actea, s/n, E-38205, La Laguna, Tenerife, Spain}
\affiliation{Departamento de Astrof\' isica, Universidad de La Laguna, E-38206, La Laguna, Tenerife, Spain}

\author{C. Ramos Almeida}
\affiliation{Instituto de Astrof\' isica de Canarias, Calle V\' ia L\'actea, s/n, E-38205, La Laguna, Tenerife, Spain}
\affiliation{Departamento de Astrof\' isica, Universidad de La Laguna, E-38206, La Laguna, Tenerife, Spain}

\author{Jean Koulidiati}
\affiliation{Laboratoire de Physique et de Chimie de l'Environnement, Universit\'e Joseph Ki-Zerbo, 03 BP 7021, Ouaga 03, Burkina Faso}




\begin{abstract}

Observed scaling relations in galaxies between baryons and dark matter global properties are key to shed light on the process of galaxy formation and on the nature of dark matter. Here, we study the scaling relation between the neutral hydrogen (H\i) and dark matter mass in isolated rotationally-supported disk galaxies at low redshift. We first show that state-of-the-art galaxy formation simulations predict that the H\i -to-dark halo mass ratio decreases with stellar mass for the most massive disk galaxies. We then infer dark matter halo masses from high-quality rotation curve data for isolated disk galaxies in the local Universe, and report on the actual {\it universality} of the H\i -to-dark halo mass ratio for these observed galaxies. This scaling relation holds for disks spanning a range of 4 orders of magnitude in stellar mass and 3 orders of magnitude in surface brightness. Accounting for the diversity of rotation curve shapes in our observational fits decreases the scatter of the H\i -to-dark halo mass ratio while keeping it constant. This finding extends the previously reported discrepancy for the stellar-to-halo mass relation of massive disk galaxies within galaxy formation simulations to the realm of neutral atomic gas. Our result reveals that isolated galaxies with regularly rotating extended H\i\ disks are surprisingly self-similar up to high masses, which hints at mass-independent self-regulation mechanisms that have yet to be fully understood.

\end{abstract}


\section{Introduction} \label{sec:intro}

The detailed study of the observed dynamics of galaxies and of the distribution of baryons within them led over the last decades to the establishment of important scaling relations linking the baryon content to the gravitational field of galaxies \citep[see, e.g.,][]{Lelli2022a}, which could improve our understanding of the galaxy formation process and possibly shed light on the nature of dark matter (DM). Most of these scaling relations either focus on the total amount of observable baryons in galaxies, such as in the baryonic Tully-Fisher relation \citep{McGaugh2000,Lelli2019,Schombert2022}, or on their stellar content, such as in the stellar-to-halo mass relation \citep[SHMR, e.g.,][]
{Moster2010, Behroozi2010,Moster2013,Posti2019b}, connecting the stellar mass of a galaxy to its DM halo mass.

The SHMR is particularly important as it indicates how much stellar mass has assembled out of the primeval amount of baryons expected in galaxies from the cosmic baryon fraction. It can also be compared to expectations from abundance matching, a technique that matches halos -- from the theoretical halo mass function expected in the standard $\Lambda$CDM cosmological model -- with observed galaxies -- following a luminosity function with a very different shape. The SHMR expected from abundance matching cannot be represented by a simple power-law and displays a turnover at high mass \citep[e.g.,][]{Moster2010, Behroozi2010}. This characteristic shape of the SHMR is nowadays a clear constraint and output for most galaxy formation simulations \citep{Schaye2015,Pillepich2018,Marasco2020}. However, resolved galaxy rotation curves have revealed that the SHMR of nearby isolated disk galaxies is actually a simple power-law with no turnover at high mass \citep{Posti2019b,Marasco2020,Posti2021}. Its linear shape on a logarithmic scale, together with other such scale-free relations (e.g., the baryonic Tully-Fisher relation and the mass-size relation), contributes to portraying isolated disks as a re-scalable population of objects, implying self-regulating mechanisms that are yet to be fully understood.

On the other hand, less attention has been paid to the relationship between the cold gas and the DM content of disk galaxies, whilst it is known that cold gas plays a crucial role in the process of star formation. Here, we therefore focus on the less-explored neutral atomic hydrogen-to-halo mass relation (H\i HMR). Similarly to the SHMR, various methods allow one to probe the H\i HMR observationally, e.g. galaxy clustering \citep{Guo2017, Padmanabhan2017a,Obuljen2019,Calette2021}, halo abundance matching \citep{Popping2015, Padmanabhan2017,Chauhan2021}, or H\i\ spectral stacking \citep{Guo2020}. However the main assumption of abundance matching, namely that there is a direct relation between the halo mass and a galaxy property, certainly does not hold for H\i\ which can be very sensitive to other factors, such as, e.g., its morphology. Recent studies, such  as that of \cite{ Dutta2022}, have for instance attempted to derive the H\i HMR for H\i -selected galaxies only, also separating the sample into blue and red galaxies. Despite these various different methods and selections, a common finding of {\it all} these studies is that the relation between H\i\ mass and halo mass at $z=0$ is typically described by a double power-law with a turnoff halo mass between $10^{11}$ and $10^{12} ~\rm M_\odot$. This typically translates into a H\i -to-halo mass ratio that varies with the galaxy halo mass or stellar mass with a peak and plateau around that characteristic mass and a decrease on both ends of the relation. However, all these determinations are slightly model-dependent as they rely on an {\it a priori} knowledge of the halo mass function, and none of them are therefore {\it direct} determinations. Here, we will thus check whether this result holds when analyzing individual observed galaxy rotation curves of isolated disk galaxies in the local Universe, and directly compare these results to the predictions of state-of-the-art galaxy formation simulations for disk galaxies.

\section{The H\i HMR in state-of-the-art simulations}
\label{simulations}

\subsection{Illustris-TNG50}

We first invetsigate the H\i HMR in the N-body/hydrodynamical simulations Illustris-TNG
\citep{Pillepich2018,Nelson2018,Nelson2019,Pillepich2019}. We  extracted from the highest resolution realisations TNG50-1 (for which the mean baryon and DM particle mass resolutions are respectively $\rm 8.5 \times 10^{4}M_{\odot}$ and $\rm 4.5 \times 10^{5}M_{\odot}$) the central subhaloes of stellar mass $\rm M_\star$ and DM halo mass $\rm M_{200}$ from the group catalog (which contains about 4.4 million subhaloes) and cross-matched them with the supplementary  H\i $\rm H_2$ catalog, computed in post-processing with the methods described in \citet{Diemer2018}, to extract the corresponding H\i\ mass ($\sim 3 \times 10^4$ subhalos). Selecting only halo masses $\rm \geq 10^9 M_{\odot}$, we end up with 21168 TNG50-1 central (and mostly isolated) galaxies at redshift $z=0$ having H\i\ mass available. Their distribution of H\i\ versus stellar mass is displayed on Fig.~\ref{fig:MstarMgas}.

\begin{figure*}
\includegraphics[width=9.3cm]{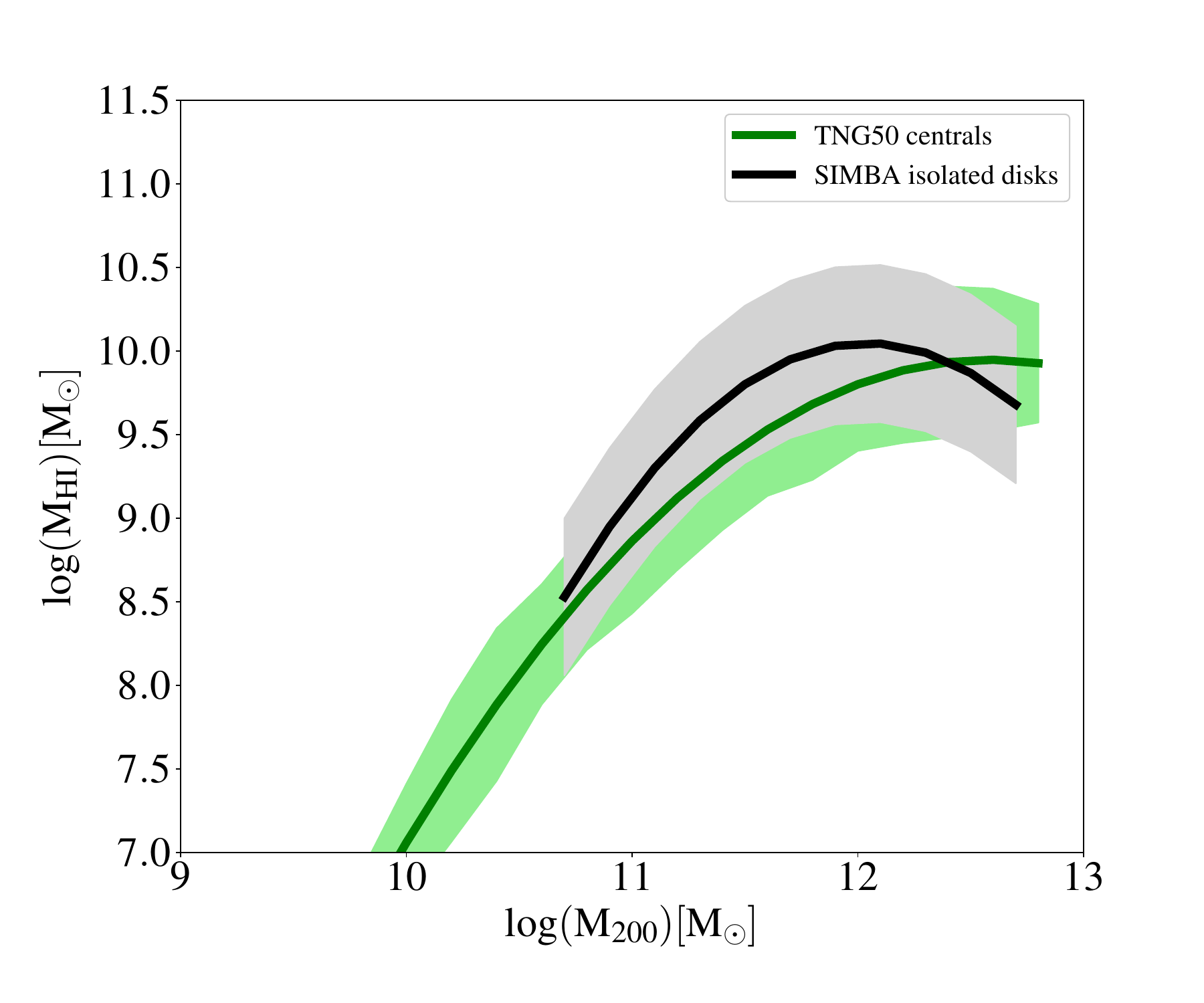}
\includegraphics[width=9.3cm]{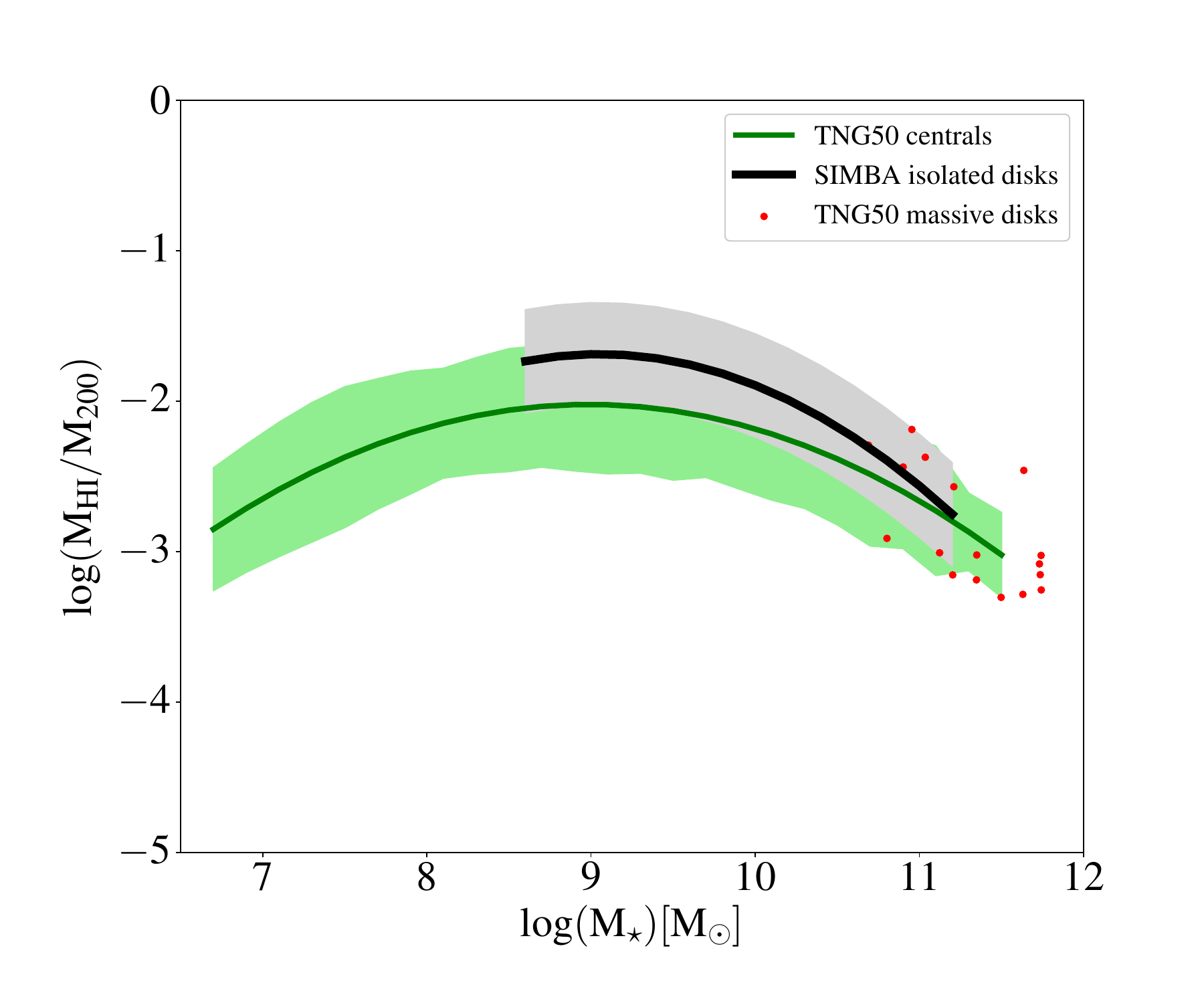}

\caption{\textbf{Left panel}: H\i -mass vs Halo mass for $\sim$ $2.1 \times 10^4$ central galaxies (irrespective of morphology) in Illustris-TNG50 (green), and for 3180 isolated {\it disk} galaxies in SIMBA with M$_{200} \geq 10^{9} M_{\odot}$ and \vsig $\geq$0.8 (grey). The lines represent the median of the binned data and the 1$\sigma$ band around the median is represented by the green and grey bands, respectively. \textbf{Right panel}: H\i -mass to Halo-mass ratio vs stellar mass. The 1$\sigma$ band around the median is represented by the green and grey bands. Green: $\sim$ $2.1 \times 10^4$ centrals in TNG50. Red dots: simulated galaxies with $\rm M_{\star} > 5\ x\   10^{10}M_{\odot}$ in TNG50 and with the fractional mass of stars with a stellar circularity $\epsilon > 0.7$ larger than 30\%, hence disk galaxies in TNG50. Grey: 3180 isolated disk galaxies in SIMBA, with M$_{200} \geq 10^{9} M_{\odot}$ and \vsig $\geq$0.8.}
\label{fig:MHMhTNG50}
\end{figure*}

Fig.\ref{fig:MHMhTNG50} (left panel, green) is constructed using these 21168 TNG50-1 centrals, showing $\rm M_{HI}$ as a function of M$_{200}$. To check whether massive disk galaxies follow the same trend as the general population at high masses, we then concentrated on the 271 central galaxies with M$_{200} \geq 10^{12} M_{\odot}$ : each of these galaxies has H\i\ mass associated to it in the supplementary H\i$\rm H_2$ catalog. We then selected disk galaxies among those, with a criterion based on having a stellar circularity parameter fraction $f(\epsilon>0.7) > 0.3$ \citep{Tacchella2019}: this stellar circularity parameter for each stellar orbit, $\epsilon$, is the ratio between the vertical component of the angular momentum and its value for a circular orbit. The right panel of Fig.\ref{fig:MHMhTNG50} shows in red dots the 18 galaxies with $\rm M_{\star} > 5\ x\   10^{10}M_{\odot}$ and a circularity parameter fraction $f(\epsilon>0.7) > 0.3$, which clearly shows that these galaxies do follow the break in $\rm M_{ HI}/M_{200}$ at high stellar masses, suggesting that this predicted break is independent of the morphological type in the TNG50 simulation. \\

\subsection{SIMBA}

While it appears that massive disk galaxies in TNG50 display a decrease of the H\i -to-dark halo mass ratio at the high-mass end, it should be noted that this conclusion is based on rather small number statistics, and most importantly that TNG50 does not resolve H\i\ gas masses, which are treated in post-processing and therefore certainly subject to some non-negligible uncertainty. It is thus desirable to also investigate the H\i HMR relation for disk galaxies in simulations that can compute H\i\ masses self-consistently on the fly. 

%
\simba\footnote{\url{http://simba.roe.ac.uk/}} \citep{Dave2019} is a suite of cosmological hydrodynamical simulations run with a modified version of the gravity and hydrodynamics solver \gizmo\ \citep{Hopkins2015}, based on the \gad\ gravity solver \citep{Springel2005}. It tracks gas particles self-consistently, on the fly, via sub-grid prescriptions to account for molecular gas production and destruction, and approximate self-shielding that results in neutral gas. 
The H$_2$ is computed using the sub-grid prescription based on the local metallicity and gas column density following \cite{KrumholzGnedin2011}. The self-shielding that results in the total neutral gas is based on the prescription of \cite{Rahmati2013}, with  the ionizing flux strength attenuated depending on the gas density and assuming a spatially uniform ionizing background from \cite{HaardtMadau2012}. Subtracting off the H$_2$ from the neutral gas then gives the H\i\ contribution.

We analyze the \simba\ run following the evolution of 1024$^3$ dark matter (DM) and 1024$^3$ gas particles within a comoving volume of (100 $h^{-1}$ Mpc$)^3$, but we checked that results are consistent if concentrating on a higher resolution 25~$h^{-1}$~Mpc volume with less statistics. To describe the morphology of galaxies, we use the kinematic ratio of their rotation- to dispersion-dominated velocity, \vsig. This quantity is computed from the 3D velocity distribution of stellar particles of each galaxy, the tangential velocity being computed in the plane perpendicular to the angular momentum of the stellar component of a galaxy.
Note that this quantity is not directly comparable to observational measures of \vsig. We simply use it to separate rotation-dominated from dispersion-dominated systems. We select the 3180 central galaxies that have M$_{200} \geq 10^{9} M_{\odot}$, \vsig $\geq$0.8, and that are isolated, which we consider as isolated disks. For the isolation criteria, we select those for which the number of galaxies in the halo is equal to 1. Their distribution of H\i\ versus stellar mass is displayed on Fig.~\ref{fig:MstarMgas}. It then appears clearly from Fig.\ref{fig:MHMhTNG50} (grey bands) that the conclusion reached from TNG is confirmed with a much higher number of identified isolated disks and a self-consistent treatment of the gas in SIMBA.

\section{Mass modeling of isolated disks}

After having analyzed the H\i HMR of disk galaxies in two modern state-of-the-art galaxy formation simulations, we now turn to a direct comparison to data. As outlined in the introduction, most methods used to measure this relation are indirect and somewhat model-dependent. 
A precise assessment of the H\i HMR in disk galaxies can actually {\it only} be achieved through individual detailed mass-modelling.

We start from a sample of 175 isolated nearby rotationally-supported galaxies, namely 158 SPARC\footnote{Spitzer Photometry and Accurate Rotation Curves} \citep{Lelli2016} and 17 LITTLE THINGS\footnote{Local Irregulars That Trace Luminosity Extremes, The H\i\ Nearby Galaxy Survey} \citep{Hunter2012,Oh2015,Iorio2017,Read2017} galaxies.
To avoid systematic uncertainties related to inclination in nearly face-on systems, only SPARC galaxies with inclinations larger than 30 degrees were kept in our analysis. All these galaxies were selected with observed extended H\i\ disks and regular disk-like kinematics. One has to keep in mind that all the observational results reported hereafter pertain to such {\it regularly rotating} disks, especially when comparing data to simulations. With this caveat in mind, we note that, in terms of $\rm M_{HI}\ vs.\ M_{\star} $, the observed galaxies analyzed here are no outliers from the general population from the blind extra-galactic H\i\ survey ALFALFA \citep{ALFALFA}, and compare well with simulated ones.

To consider the contribution of the gas to the circular velocity for the LITTLE THINGS galaxies, we derived the expected circular velocity curve from the H\i\ surface densities taken from \citet{Iorio2017} for which we applied a multiplicative factor of 1.33 to account for the presence of Helium. Since these galaxies are very gas-rich, the stellar mass-to-light ratios from their B-band photometry are less important to know precisely than for most SPARC galaxies. For the latter, 3.6 $\mu$m photometry allows to minimize the variations of the stellar mass-to-light ratios from stellar population synthesis models \citep{Meidt2014,McGaugh2014,Schombert2019}. 

To investigate the relation between neutral hydrogen and DM mass within the sample, we then produced mass models with two analytical DM profiles: (i) the standard Navarro-Frenk-White \citep[{\rm NFW}][]{Navarro1996} cuspy density profile, characterised by a dimensionless concentration parameter (c) and the halo mass (M$_{200}$), and (ii) the so-called Dekel-Zhao \citep[{\rm hereafter DZ; }][]{Zhao1996, Dekel2017, Freundlich2020b} profile, characterised by a variable inner slope ($\rm s_1$) defined as the absolute value of the logarithmic slope at 1\% of the virial radius, a dimensionless concentration parameter (c) and the halo mass (M$_{200}$). We assume a Gaussian prior for the logarithm of the mass-to-light ratio of the stellar disc, centred on $\rm M/L_{disk}=0.6\,\rm M_\odot/L_\odot$ with a dispersion $\rm \sigma$ = 0.2 dex for SPARC data and centred on 1 $\rm M_\odot/L_\odot$ with $\rm \sigma$ = 0.2 dex for the B-band LITTLE THINGS data, which broadly encompasses the values expected from stellar population synthesis models \cite{Meidt2014,McGaugh2014}. For the halo concentration $c$, we assume a Gaussian prior that follows the $c$-$\rm M_{200}$ relation as estimated in \cite{Dutton2014} with a scatter of 0.11. For the DM halo mass M$_{200}$, we use a flat prior over a wide range of $\rm 0 \leq {\rm log} M_{200}/M_{\odot} \leq 20$. For galaxies with a spherical bulge component, we assume that $\rm M/L_{bulge} = 1.4 \times M/L_{disk}$, as suggested by stellar population synthesis models \cite{Schombert2014}. For the additional free parameter $\rm s_1$ of the DZ profile, we assume a flat prior in the range $0 \leq s_1 \leq 5$.

The parameters are then fitted using an affine-invariant Markov chain Monte Carlo (MCMC) sampling with the python implementation EMCEE \cite{Foreman-Mackey2013}. For each parameter, we take the median of the marginalised posterior as the best-fit value while lower and upper errors are taken at the 16th and 84th percentiles. For 25 galaxies, typically with $\rm {\rm log}(M_{200}) <  9$, the fits are of very bad quality in both cases, and we left these galaxies out of our analysis, ending up with a sample of 150 galaxies for both profiles NFW and DZ.

 The reduced $\chi^2$ values for all NFW and DZ fits are given in Table \ref{tab:my_label}. The median reduced $\chi^2$ is 1.93 for NFW (with a large dispersion of 5.3, toward very bad fits) and 1.05 for DZ (with a dispersion of 2.8). An example is given on Fig.~\ref{fig:NFW and DZ mass models}. In the following, we concentrate on the DZ profile which gives significantly better fits, taking into account the full diversity of rotation curve shapes. The median stellar mass-to-light ratio for the DZ fits of the SPARC galaxies is $\rm M/L_{disk}=0.53\,\rm M_\odot/L_\odot$. The median is closer to the prior for NFW fits (0.57), but it is in this case boosted by some high values from poor NFW fits.

\section{Results}
While we concentrate hereafter on the results obtained from the DZ fits, most of the scaling relations reported hereafter are unaltered when using NFW profiles, with one exception (the independence on surface brightness) but the {\it scatter} around the scaling relations typically increases when using NFW fits, meaning that bettrer individual fits lead to tighter scaling relations. We have also checked that our results hold for all parametrizations of the DM halos of SPARC galaxies used by \cite{Li2020}.\\

\begin{figure*}
\includegraphics[width=9.3cm]{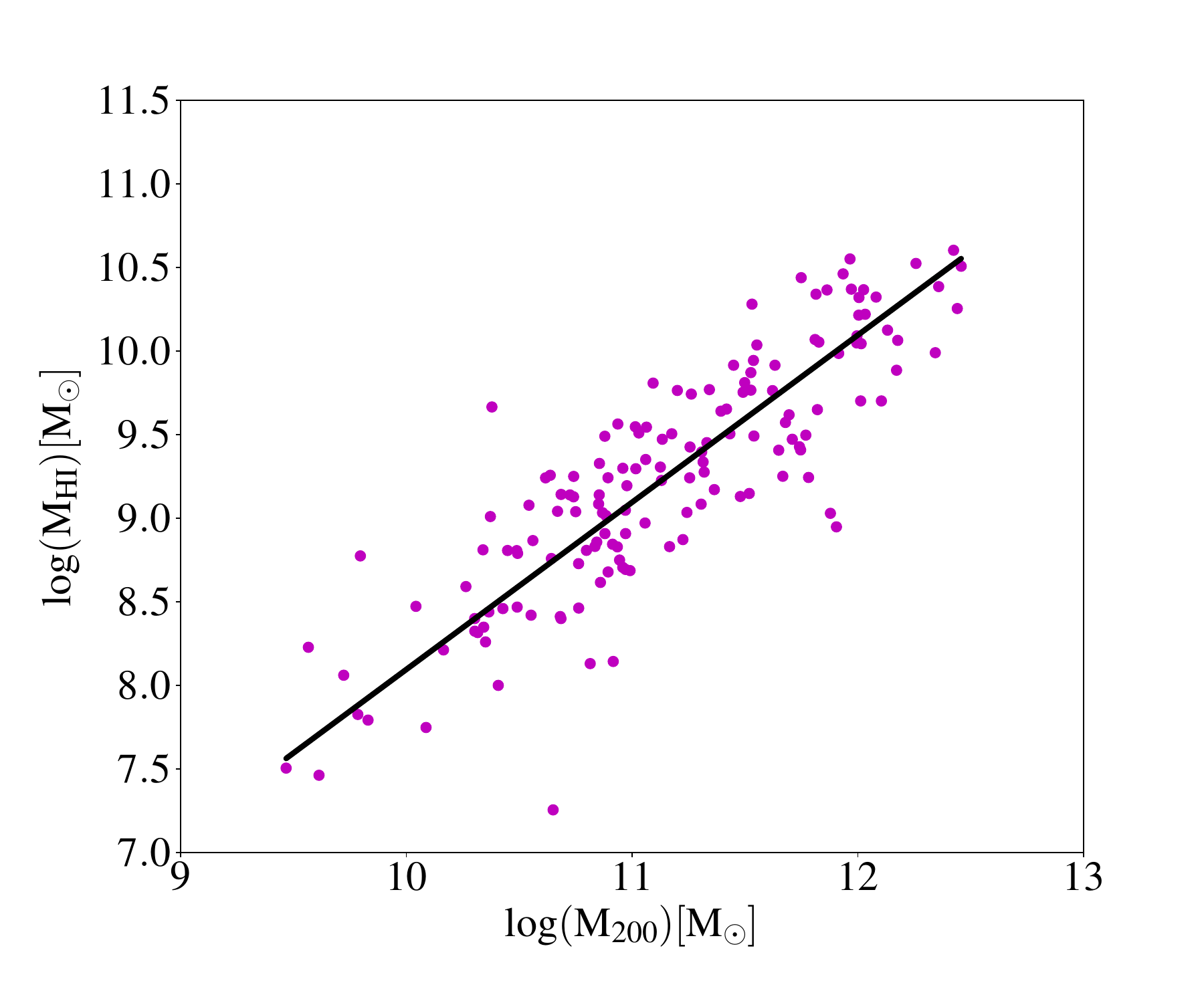}
\includegraphics[width=9.3cm]{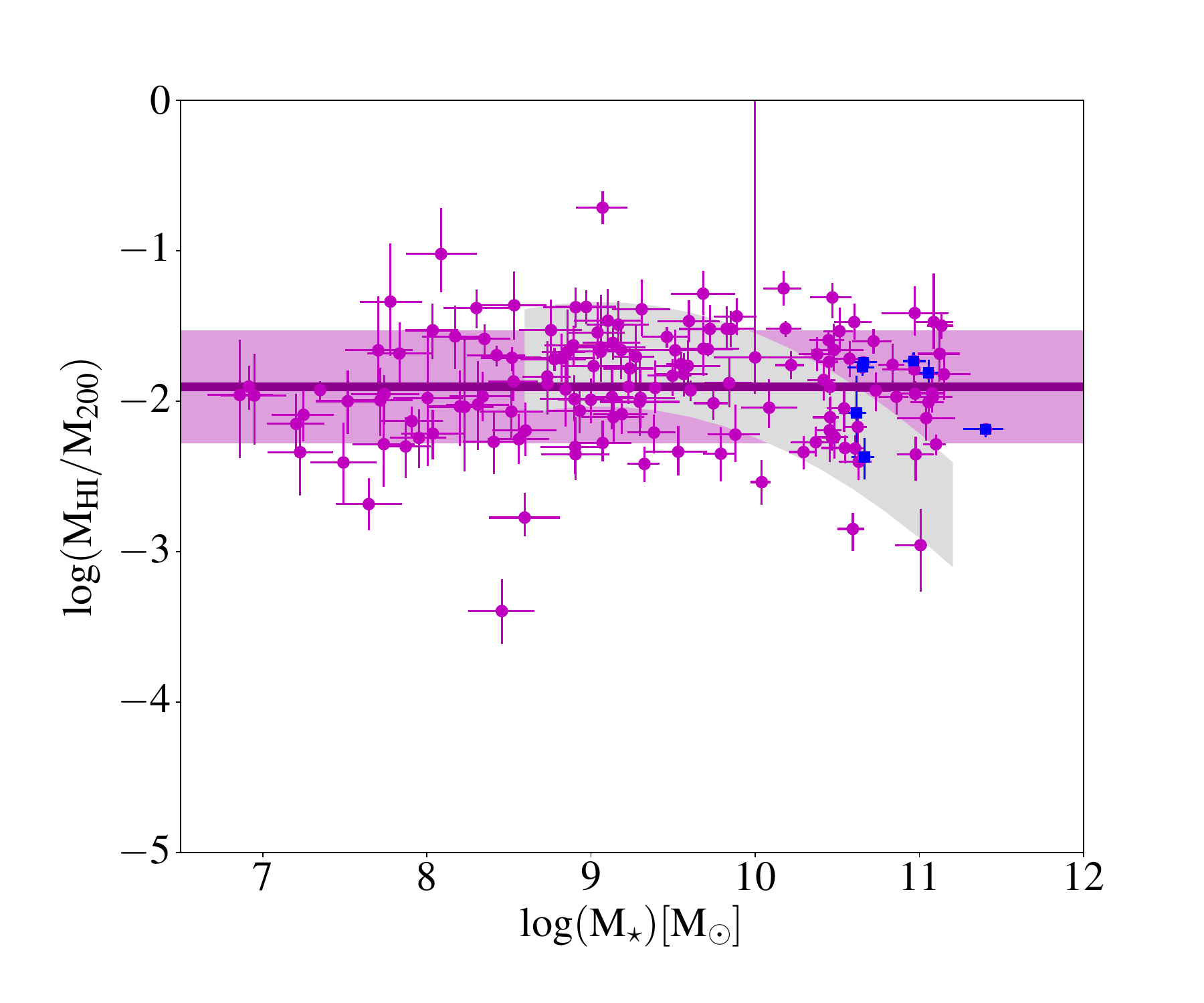}\\
\includegraphics[width=9.3cm]{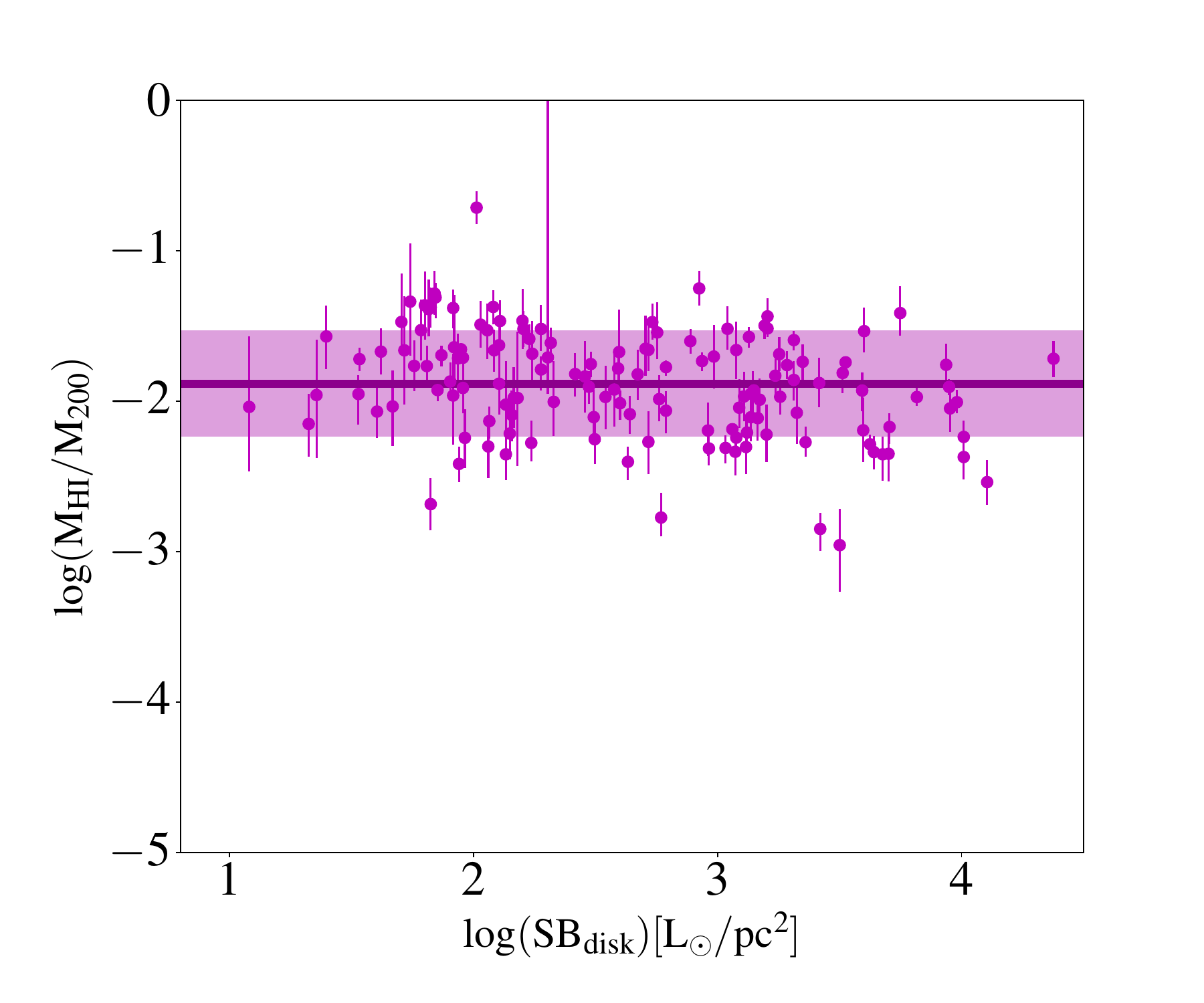}
\includegraphics[width=9.3cm]{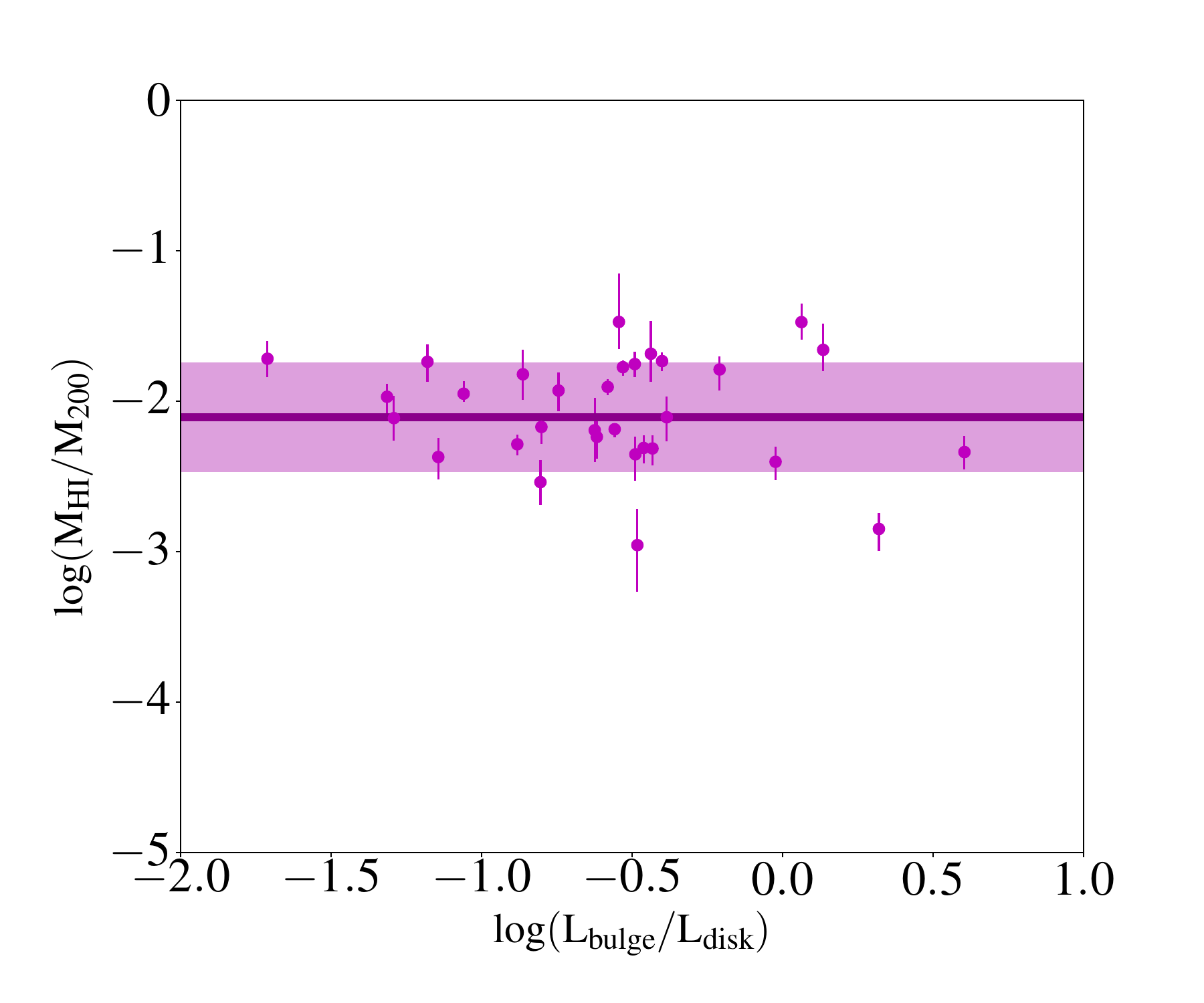}
\caption{\textbf{Top left}: H\i-mass vs Halo mass for the 150 rotation curve fits with DZ profiles (magenta dots). The black line corresponds to the relation $ \rm M_{HI} = log(median(M_{HI}/M_{200})) \times M_{200}$. \textbf{Top right}: H\i -to-Halo-mass ratio vs stellar mass for the same sample (magenta dots with their associated error bars) : the magenta line corresponds to the median value and the band to the 1 $\sigma$ zone (scatter = 0.37 dex; intrinsic scatter = 0.31 dex). Blue squares depict galaxies hosting an AGN. The grey band indicates the 1$\sigma$ band from isolated disks in SIMBA as a comparison.
\textbf{Bottom left}: H\i-to-Halo-mass ratio for the DZ fits of the SPARC galaxies as a function of disk central surface brightness at 3.6 $\mu$m. \textbf{Bottom right}: Same as a function of bulge to disk luminosity ratio for 31 galaxies with bulge.}
\label{fig:MHIM200Ms}
\end{figure*}

The main results of our analysis are all summarized on Fig.\ref{fig:MHIM200Ms}. The H\i\ mass as a function of M$_{200}$ of the 150 galaxies of the sample displays a linear correlation between the two parameters, without any sign of a break at high mass. This implies that the H\i\ mass is linearly increasing with halo mass for isolated disk galaxies in the local Universe. In other words, the H\i -mass to Halo-mass ratio of disk galaxies appears to be compatible with being constant. 
To check that this finding is independent of stellar mass, we display the ratio $\rm M_{HI}/M_{200}$ as a function of stellar mass, with the error-bars as estimated from the DZ profile parametrization. The ratio is indeed compatible with a constant value of 1.25\% (i.e., ${\rm log \, (M_{HI}/M_{200})}= -1.903$) and an intrinsic scatter of 0.31~dex, namely a factor of 2, which is remarkable given that the sample spans 4 orders of magnitude in stellar mass. Note that this ratio does remain constant with stellar mass when using the NFW parametrization too, albeit with a higher intrinsic scatter of 0.37~dex.  Therefore, accounting for the actual diversity of rotation curve shapes (and DM profile cores and cusps) in our observational fits decreases the scatter of the ratio while keeping the ratio itself constant, which strengthens the robustness of the result. This is to be contrasted with the results from state-of-the-art simulations where the ratio decreases significantly at high masses. As a caveat, we note that the precise selection function to get the exact same population in simulations is not known, but we checked that the simulated disk galaxies having the {\it highest} $\rm M_{HI}/M_{200}$ with $\rm M_{\star}/M_{\odot} > 10^{11}$ fall below the median of the observations (Fig.~\ref{fig:MstarMgas}).

To assess the robustness of our finding on this relatively small sample, we performed a Kolmogorov-Smirnov test on $\rm M_{HI}/M_{200}$ based on stellar mass using our sample of 150 galaxies. The test was run to compare the cumulative distributions of two data sets:  galaxies with $10^8 \leq {\rm M}_{\star}/{\rm M}_{\odot} < 10^{10}$ and galaxies with ${\rm M}_{\star} \geq 10^{10} {\rm M}_{\odot}$. We found a statistic of 0.194 and a p-value of 0.154, hence the null hypothesis that the two samples are indeed drawn from the same distribution cannot be rejected. Although not statistically significant based on our sample size, we nevertheless note that when binning the data, a small downward trend seems to appear at the lowest mass end (Fig.~\ref{fig:MstarMgas}). Interestingly, we find that, at the high mass end, massive disks hosting active galactic nuclei (AGN) identified based on their X-ray emission (e.g., from XMM-Newton and Chandra) follow exactly the same relation as the other galaxies (Fig.\ref{fig:MHIM200Ms}).
We also checked that the ratio remained constant over $\sim$3 orders of magnitude in surface brightness in DZ fits (note that NFW fits did lead to a dependency here, but associated to the fact that low surface brightness galaxies have a higher tendency to be actually cored), and that it did not depend on whether the galaxy hosts a bulge (which corresponds to 31 galaxies in total).

\section{Conclusion}

In this Letter, we report on the unexpected universality of the $\rm M_{HI}/M_{200}$ ratio for isolated disk galaxies with extended H\i\ rotation curves in the local Universe. From our study, there appears to be no correlation between the halo mass, stellar mass, or surface brightness of disk galaxies and their $\rm M_{HI}/M_{200}$ ratio, which remains remarkably universal at a value of $\sim 1.25$\%, within a factor of 2 at 1~$\sigma$. While it has been known for a long time that both stellar mass and total baryonic mass vary strongly across the disk galaxy population, it appears that this is not the case for H\i\ in isolated disks. That the H\i\ mass of an isolated disk galaxy is a direct tracer of its halo mass appears counter-intuitive, and could hint at interesting self-regulating mechanisms. Studying how this $\rm M_{HI}/M_{200}$ ratio varies when quenching takes place to transform star-forming galaxies into gas-poor red and dead ones would be an interesting follow-up to understand both the self-regulation mechanisms at play and the dominant quenching mechanisms. 

It had already been shown that the SHMR of isolated disk galaxies appears to be a simple power law with no turnover at high masses, indicating that the fundamental parameters of disk galaxies may be single-slope monotonic functions of mass, with a small scatter, instead of being complicated non-monotonic functions \citep{Posti2019b}. The present finding confirms this picture in a spectacular fashion: the $\rm M_{HI}/M_{200}$ ratio of disks does not even depend on mass. 

A corollary of the single power-law SHMR for isolated disks is that massive disks are actually too dark matter dominated within modern state-of-the-art simulations of galaxy formation \citep{Marasco2020}. In other words, the simulated SHMR does not vary with disc fraction and Hubble type as it should from observations \citep{Posti2021}, and AGN feedback does not seem to work as expected from simulations in observed disk galaxies, i.e. not expelling as many baryons from massive halos as expected, a problem known as the ``failed feedback" problem. Interestingly, AGN-hosting galaxies follow exactly the same trend as other galaxies in our observed sample, supporting the ``failed feedback" interpretation of this discrepancy.

The discrepancy between our findings and the simulations TNG50 and SIMBA, where the $\rm M_{HI}/M_{200}$ ratio decreases with mass for massive disks, is therefore yet another manifestation of the fact that simulated massive disks are too dark matter dominated indeed. 

We note that alternative frameworks having the baryonic Tully-Fisher built-in as a fundamental relation \citep{Milgrom1983,Famaey2012} do predict a monotonically rising SHMR and baryonic-to-halo mass relation (BHMR), but the universality of the H\i HMR reported here does not follow naturally from there, as it results from a conspiracy between the BHMR (following from the baryonic Tully-Fisher relation) and the scaling relation between the stellar and gas mass of galaxies \citep{Oria2021}, so as to yield a universal $\rm M_{HI}/M_{200}$ ratio.

In conclusion, this universal ratio points to isolated rotationally-supported star-forming disk galaxies of all masses and surface densities being surprisingly self-similar, which hints at mass-independent self-regulation mechanisms that are yet to be fully understood. 

\section*{Acknowledgments}
The authors thank the anonymous referee for a thoughtful report that helped improve the manuscript. They thank Dylan Nelson for providing detailed information on the Illustris-TNG hydrodynamical simulations. The authors also thank Romeel Dav\'e for making the hydrodynamical simulation Simba publicly available. They are greatful to Natasha Maddox for useful discussions.
M.K. acknowledges funding from the Schlumberger Foundation Faculty for the Future program, and from the Observatoire Astronomique de Strasbourg (ObAS). M.K. knowledges financial support from the Women by Science Programme of the Fundaci\'on Mujeres por \'Africa, from the Gobierno de Canarias, and from the Instituto de Astrof\'isica de Canarias (IAC). B.F., J.F., L.P. and R.I. acknowledge funding from the European Research Council (ERC) under the European Unions Horizon 2020 research and innovation program (grant agreement No. 834148) and from the Agence Nationale de la Recherche (ANR projects ANR-18-CE31-0006 and ANR-19-CE31-0017). CRA acknowledges the project “Feeding and feedback in active galaxies”, with reference PID2019-106027GB-C42, funded by MICINN-AEI/10.13039/501100011033.

\bibliography{sample631}{}
\bibliographystyle{aasjournal}

\appendix
\restartappendixnumbering

\section{Stellar vs. gas masses and binned data}

\begin{figure*}[ht]
\includegraphics[width=9.3cm]
{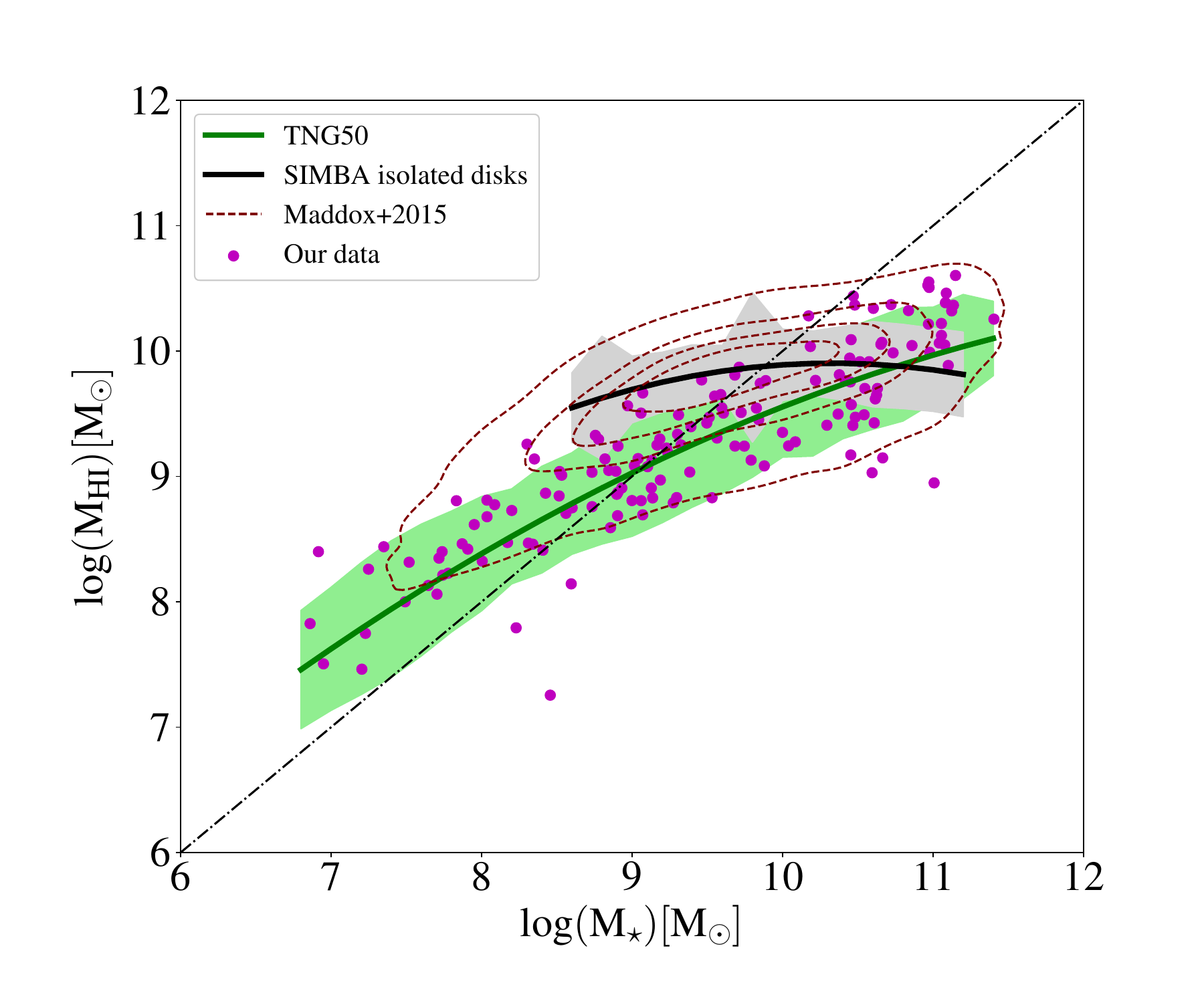}
\includegraphics[width=9.3cm]
{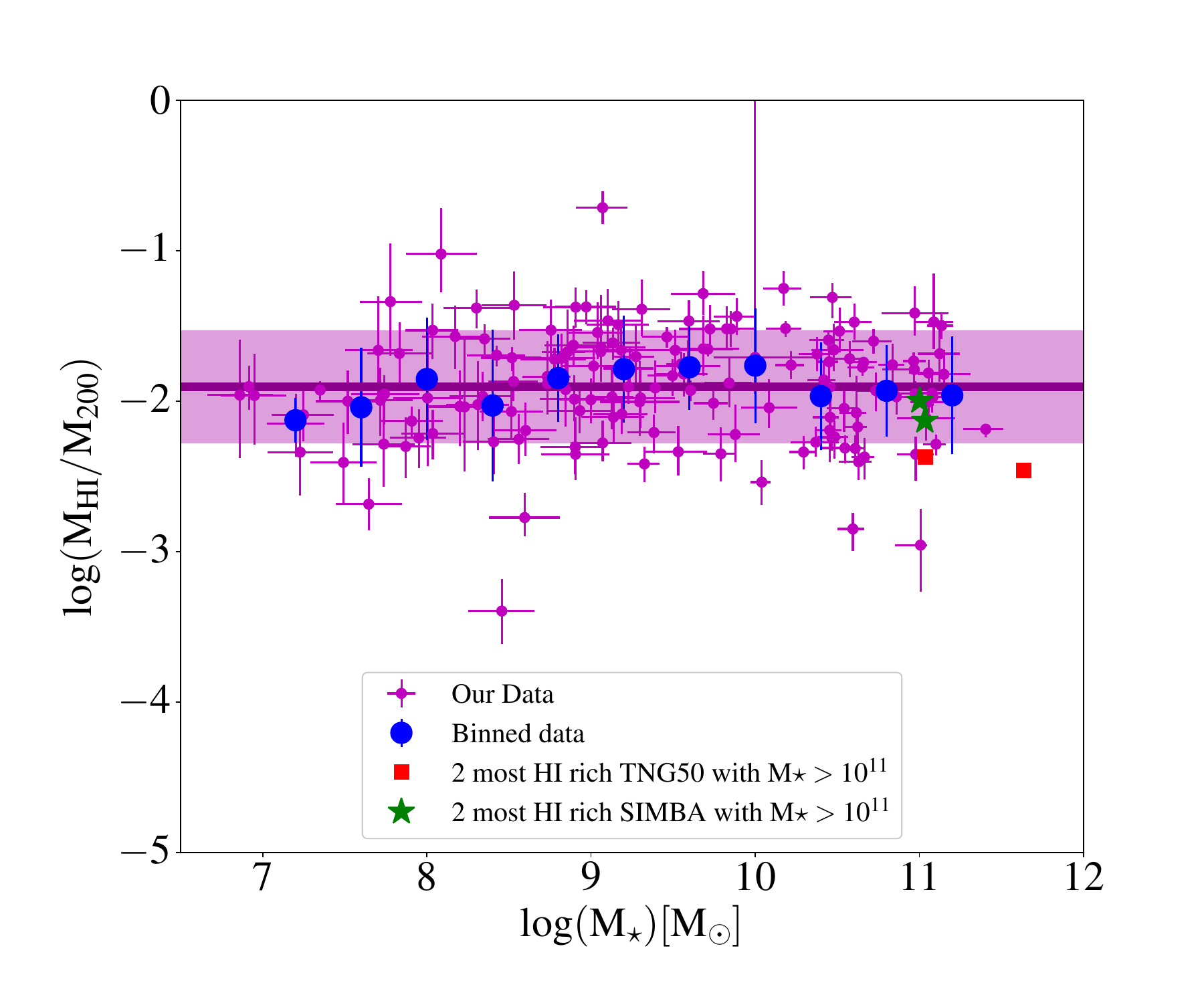}
\caption{\textbf{Left panel}: H\i\ mass as a function of stellar mass.  The individual measurements for the 150 galaxies are presented in magenta dots. The contours correspond to the 9153 H\i -selected ALFALFA galaxies from \citep[][see that paper for details]{Maddox2015}. The green and grey lines (bands) represent the median (1$\sigma$ scatter) of the binned data of central galaxies in Illustris-TNG50 and isolated disk galaxies in SIMBA respectively. The diagonal dashed line marks the one-to-one relation of equal H\i\ and stellar masses. \textbf{Right panel}: H\i-to-Halo-mass ratio vs stellar mass for the 150 galaxies with blue dots indicating the median and scatter in each bin. The two simulated disk galaxies in both Illustris-TNG50 and SIMBA having $\rm M_{\star}/M_{\odot} > 10^{11}$ and the {\it highest} $\rm M_{HI}/M_{200}$ are represented in red squares and green stars respectively.}
\label{fig:MstarMgas}
\end{figure*}

To broadly compare the subpopulation of galaxies studied here with the general population and with the simulations, we show on Fig.\ref{fig:MstarMgas} the distribution of gas vs. stellar mass of the TNG50 centrals, the SIMBA disks and the observed galaxies analyzed here, together with isocontours from the general population of the blind extra-galactic H\i\ survey ALFALFA.  One has however to keep in mind that all the observational results reported in this Letter pertain to regularly rotating extended H\i\ disks. For instance, if observed earlier type galaxies were included at the high mass end, the observational data set would naturally also turn down to some degree at the high mass end. Also, since the simulated data is binned in the rest of the paper, we display on Fig.\ref{fig:MstarMgas} our main result when the observed data is binned, showing a slight downward trend at the low mass end. Finally, Fig.\ref{fig:MstarMgas} also displays the two simulated disk galaxies  in both TNG50 and SIMBA having the highest $\rm M_{HI}/M_{200}$ with $\rm M_{\star}/M_{\odot} > 10^{11}$ : since those still fall below the median of the observations, it confirms a tension independently of the above caveats.

\section{Rotation curve fits}

\centering
\begin{longtable}{l c c c c c c c c c c c c c c c}
\caption{Properties of the sample of 150 galaxies studied in this work: (1) name of the galaxy; (2) and (5) stellar mass in  $\rm M_\odot$ fron NfW and DZ respectively with their 16th-84th percentiles; (8) total H\i\ mass in $\rm M_\odot$ from \cite{Iorio2017}; (9) and (12) dark matter halo mass $\rm M_{200}$ in $\rm M_\odot$ derived from NFW and DZ respectively with their 16th-84th percentiles; (15) and (16) reduced $\chi^2$ from NFW and DZ respectively.}
\label{tab:my_label}\\

\hline\hline                 
\multicolumn{1}{l}{\textbf{$\rm Galaxy$}} & \textbf{$\rm M_{{\star}_{NFW}}$} & \textbf{$\rm16th$} & \textbf{$\rm84th$} & \textbf{$\rm M_{{\star}_{DZ}}$} & \textbf{$\rm 16th$} & \textbf{$\rm84th$} & \textbf{$\rm M_{HI}$}  & \textbf{$\rm M_{{h}_{NFW}}$} & \textbf{$\rm 16th$} & \textbf{$\rm84th$} & \textbf{$\rm M_{{h}_{DZ}}$} & \textbf{$\rm16th$} & \textbf{$\rm84th$} & \textbf{$\rm \chi^2_{r_{NFW}}$} & \textbf{$\rm \chi^2_{r_{DZ}}$}\\    
\multicolumn{1}{l}{(1)} & {(2)} & {(3)} & {(4)} &{(5)} &{(6)} & {(7)} & {(8)} & {(9)} & {(10)} & {(11)} & {(12)} & {(13)} & {(14)} & {(15)} & {(16)} \\

\hline
\endfirsthead

\multicolumn{16}{l}%
{\tablename\ \thetable{} -- {\it continued from previous page}} \\ 
\hline
\multicolumn{1}{l}{\textbf{$\rm Galaxy$}} & \textbf{$\rm M_{{\star}_{NFW}}$} & \textbf{$\rm16th$} & \textbf{$\rm84th$} & \textbf{$\rm M_{{\star}_{DZ}}$}   & \textbf{$\rm16th$} & \textbf{$\rm84th$} & \textbf{$\rm M_{HI}$}  & \textbf{$\rm M_{{h}_{NFW}}$} & \textbf{$\rm 16th$} & \textbf{$\rm84th$} & \textbf{$\rm M_{{h}_{DZ}}$} & \textbf{$\rm16th$} & \textbf{$\rm84th$} & \textbf{$\rm \chi^2_{r_{NFW}}$} & \textbf{$\rm \chi^2_{r_{DZ}}$}\\
\multicolumn{1}{l}{(1)} & {(2)} & {(3)} & {(4)} &{(5)} &{(6)} & {(7)} & {(8)} & {(9)} & {(10)} & {(11)} & {(12)} & {(13)} & {(14)} & {(15)} & {(16)} \\ 
\hline 
\endhead

\hline \multicolumn{16}{l}{{\it Continued on next page...}} \\
\endfoot

\hline \hline
\endlastfoot

     D564-8    &  7.19	   &	 7.01	&   7.36	& 	   7.20	&      7.02		&   7.37	& 	  7.46     &     9.36       &     9.14  &    9.60	 &      9.61        & 	   9.41	 &    9.83    &   5.26      &    0.63    \\	
     D631-7    &  7.84	   &	 7.67	&   7.99	& 	   7.87	&      7.70		&   8.02	& 	  8.46     &    10.84       &    10.60  &   11.10	 &     10.76        & 	  10.56	 &   10.97    &   9.86      &    2.22    \\	
     DDO064    &  7.97	   &	 7.77	&   8.18	& 	   8.00	&      7.79		&   8.22	& 	  8.32     &    10.17       &     9.67  &   10.71	 &     10.30        & 	   9.86	 &   10.75    &   1.14      &    0.46    \\	
     DDO154    &  7.19	   &	 7.04	&   7.33	& 	   7.35	&      7.19		&   7.50	& 	  8.43     &    10.77       &    10.66  &   10.89	 &     10.36        & 	  10.30	 &   10.43    &  11.52      &    2.51    \\	
     DDO161    &  8.31	   &	 8.15	&   8.47	& 	   8.35	&      8.18		&   8.50	& 	  9.13     &    10.83       &    10.71  &   10.97	 &     10.72        & 	  10.62	 &   10.83    &   3.76      &    2.21    \\	
     DDO168    &  7.91	   &	 7.73	&   8.07	& 	   7.95	&      7.77		&   8.11	& 	  8.61     &    11.08       &    10.80  &   11.41	 &     10.85        & 	  10.67	 &   11.05    &  14.91      &    4.94    \\	
     DDO170    &  8.41	   &	 8.23	&   8.58	& 	   8.42	&      8.24		&   8.59	& 	  8.86     &    10.64       &    10.56  &   10.73	 &     10.56        & 	  10.49	 &   10.63    &   3.07      &    2.36    \\	
ESO079-G014    & 10.43	   &	10.28	&  10.56	& 	  10.36	&     10.21		&  10.51	& 	  9.49     &    12.31       &    12.08  &   12.58	 &     11.76        & 	  11.66	 &   11.86    &   4.40      &    1.70    \\	
ESO116-G012    &  9.32	   &	 9.18	&   9.44	& 	   9.38	&      9.22		&   9.51	& 	  9.03     &    11.72       &    11.52  &   11.99	 &     11.24        & 	  11.12	 &   11.37    &   2.58      &    1.18    \\	
ESO444-G084    &  7.60	   &	 7.41	&   7.79	& 	   7.64	&      7.44		&   7.84	& 	  8.13     &    11.18       &    10.92  &   11.53	 &     10.81        & 	  10.64	 &   10.98    &   0.89      &    0.17    \\	
ESO563-G021    & 11.38	   &	11.32	&  11.43	& 	  11.08	&     10.94		&  11.20	& 	 10.38     &    12.88       &    12.69  &   13.11	 &     12.35        & 	  12.29	 &   12.41    &  18.65      &    7.71    \\	
     F563-1    &  9.05	   &	 8.86	&   9.25	& 	   9.05	&      8.86		&   9.25	& 	  9.50     &    11.41       &    11.19  &   11.66	 &     11.17        & 	  11.01	 &   11.32    &   1.19      &    0.52    \\	
    F563-V2    &  9.28	   &	 9.08	&   9.49	& 	   9.30	&      9.09		&   9.51	& 	  9.33     &    11.91       &    11.48  &   12.41	 &     11.31        & 	  11.10	 &   11.51    &   1.36      &    0.43    \\	
    F565-V2    &  8.51	   &	 8.32	&   8.71	& 	   8.51	&      8.32		&   8.70	& 	  8.84     &    11.09       &    10.85  &   11.38	 &     10.91        & 	  10.75	 &   11.08    &   1.39      &    0.44    \\	
     F568-1    &  9.58	   &	 9.38	&   9.77	& 	   9.58	&      9.38		&   9.78	& 	  9.65     &    12.01       &    11.65  &   12.45	 &     11.41        & 	  11.24	 &   11.58    &   0.99      &    0.14    \\	
    F568-V1    &  9.40	   &	 9.20	&   9.61	& 	   9.39	&      9.18		&   9.59	& 	  9.39     &    11.67       &    11.36  &   12.07	 &     11.30        & 	  11.12	 &   11.47    &   0.53      &    0.18    \\	
     F571-8    &  9.22	   &	 9.10	&   9.33	& 	   9.32	&      9.22		&   9.41	& 	  9.25     &    12.40       &    12.14  &   12.72	 &     11.66        & 	  11.55	 &   11.78    &   5.34      &    1.10    \\	
    F571-V1    &  9.01	   &	 8.82	&   9.20	& 	   9.01	&      8.82		&   9.20	& 	  9.08     &    10.97       &    10.79  &   11.18	 &     10.85        & 	  10.71	 &   10.99    &   0.91      &    0.49    \\	
     F574-1    &  9.59	   &	 9.40	&   9.78	& 	   9.59	&      9.40		&   9.78	& 	  9.54     &    11.29       &    11.08  &   11.52	 &     11.01        & 	  10.87	 &   11.15    &   1.83      &    0.17    \\	
    F579-V1    &  9.97	   &	 9.75	&  10.20	& 	   9.99	&      9.74		&  10.38	& 	  9.35     &    11.25       &    10.82  &   11.59	 &     11.06        & 	   8.52	 &   11.30    &   1.13      &    0.37    \\	
     F583-1    &  8.75	   &	 8.55	&   8.94	& 	   8.75	&      8.56		&   8.94	& 	  9.32     &    11.00       &    10.73  &   11.30	 &     10.85        & 	  10.65	 &   11.05    &   2.41      &    0.46    \\	
     F583-4    &  9.03	   &	 8.82	&   9.26	& 	   9.06	&      8.84		&   9.33	& 	  8.80     &    10.56       &    10.21  &   10.87	 &     10.44        & 	  10.09	 &   10.67    &   0.61      &    0.37    \\	
     IC2574    &  8.59	   &	 8.42	&   8.74	& 	   8.52	&      8.37		&   8.67	& 	  9.01     &    11.31       &    11.13  &   11.51	 &     10.88        & 	  10.75	 &   11.01    &  43.56      &    4.96    \\	
     IC4202    & 10.61	   &	10.48	&  10.73	& 	  10.45	&     10.33		&  10.56	& 	 10.09     &    12.38       &    12.28  &   12.50	 &     11.99        & 	  11.94	 &   12.04    &  26.90      &    7.30    \\	
   KK98-251    &  7.88	   &	 7.60	&   8.41	& 	   7.70	&      7.50		&   7.91	& 	  8.06     &     9.09       &     4.64  &    9.55	 &      9.72        & 	   9.36	 &   10.08    &   5.52      &    1.20    \\	
    NGC0024    &  9.62	   &	 9.46	&   9.74	& 	   9.53	&      9.32		&   9.70	& 	  8.83     &    11.33       &    11.12  &   11.59	 &     11.16        & 	  10.99	 &   11.32    &   0.89      &    0.84    \\	
    NGC0055    &  9.20	   &	 9.04	&   9.35	& 	   9.23	&      9.07		&   9.38	& 	  9.19     &    11.18       &    10.98  &   11.42	 &     10.97        & 	  10.83	 &   11.12    &   4.66      &    1.39    \\	
    NGC0100    &  9.16	   &	 8.99	&   9.31	& 	   9.18	&      9.01		&   9.33	& 	  9.29     &    11.24       &    10.96  &   11.57	 &     10.95        & 	  10.76	 &   11.15    &   1.76      &    0.18    \\	
    NGC0247    &  9.62	   &	 9.44	&   9.79	& 	   9.68	&      9.47		&   9.90	& 	  9.24     &    11.32       &    11.09  &   11.55	 &     10.89        & 	  10.67	 &   11.07    &   2.12      &    2.09    \\	
    NGC0289    & 10.61	   &	10.50	&  10.70	& 	  10.47	&     10.33		&  10.58	& 	 10.43     &    11.84       &    11.75  &   11.94	 &     11.74        & 	  11.65	 &   11.88    &   1.94      &    2.45    \\	
    NGC0300    &  9.16	   &	 8.99	&   9.31	& 	   9.18	&      9.01		&   9.34	& 	  8.97     &    11.36       &    11.17  &   11.58	 &     11.05        & 	  10.93	 &   11.18    &   0.81      &    0.52    \\	
    NGC0801    & 11.23	   &	11.20	&  11.26	& 	  11.13	&     11.04		&  11.20	& 	 10.36     &    11.98       &    11.89  &   12.09	 &     11.86        & 	  11.77	 &   11.95    &   7.45      &   10.63    \\	
    NGC0891    & 10.69	   &	10.64	&  10.73	& 	  10.62	&     10.54		&  10.68	& 	  9.65     &    12.25       &    12.06  &   12.51	 &     11.82        & 	  11.73	 &   11.93    &   5.67      &    3.54    \\	
    NGC1003    &  9.49	   &	 9.35	&   9.61	& 	   9.46	&      9.31		&   9.59	& 	  9.76     &    11.48       &    11.39  &   11.58	 &     11.34        & 	  11.27	 &   11.41    &   3.18      &    4.83    \\	
    NGC1090    & 10.55	   &	10.46	&  10.62	& 	  10.44	&     10.33		&  10.54	& 	  9.94     &    11.73       &    11.64  &   11.83	 &     11.53        & 	  11.47	 &   11.60    &   2.51      &    1.31    \\	
    NGC1705    &  8.77	   &	 8.58	&   8.92	& 	   8.59	&      8.37		&   8.81	& 	  8.14     &    10.99       &    10.70  &   11.40	 &     10.91        & 	  10.75	 &   11.04    &   0.94      &    0.14    \\	
    NGC2366    &  7.96	   &	 7.80	&   8.12	& 	   8.03	&      7.86		&   8.18	& 	  8.81     &    10.50       &    10.28  &   10.76	 &     10.33        & 	  10.16	 &   10.53    &   4.83      &    1.05    \\	
    NGC2403    &  9.67	   &	 9.58	&   9.74	& 	   9.60	&      9.48		&   9.74	& 	  9.50     &    11.44       &    11.37  &   11.52	 &     11.43        & 	  11.36	 &   11.50    &   9.92      &    9.48    \\	
    NGC2683    & 10.73	   &	10.67	&  10.77	& 	  10.66	&     10.58		&  10.72	& 	  9.14     &    11.65       &    11.49  &   11.82	 &     11.51        & 	  11.39	 &   11.66    &   1.34      &    0.99    \\	
    NGC2841    & 11.24	   &	11.20	&  11.28	& 	  10.97	&     10.86		&  11.08	& 	  9.99     &    12.55       &    12.44  &   12.67	 &     12.34        & 	  12.22	 &   12.51    &   1.82      &    2.32    \\	
    NGC2903    & 10.51	   &	10.46	&  10.55	& 	  10.46	&     10.40		&  10.52	& 	  9.40     &    11.82       &    11.74  &   11.92	 &     11.64        & 	  11.57	 &   11.73    &   8.95      &    6.00    \\	
    NGC2915    &  8.56	   &	 8.37	&   8.74	& 	   8.56	&      8.37		&   8.74	& 	  8.70     &    11.14       &    10.91  &   11.42	 &     10.95        & 	  10.79	 &   11.12    &   1.24      &    0.97    \\	
    NGC2955    & 11.21	   &	11.18	&  11.24	& 	  11.08	&     10.97		&  11.20	& 	 10.46     &    12.13       &    11.80  &   12.45	 &     11.93        & 	  11.61	 &   12.11    &   4.80      &    2.76    \\	
    NGC2998    & 10.96	   &	10.87	&  11.03	& 	  10.72	&     10.59		&  10.83	& 	 10.37     &    12.02       &    11.93  &   12.13	 &     11.97        & 	  11.88	 &   12.05    &   2.74      &    1.10    \\	
    NGC3109    &  8.00	   &	 7.81	&   8.19	& 	   8.03	&      7.84		&   8.22	& 	  8.67     &    11.37       &    11.11  &   11.65	 &     10.89        & 	  10.73	 &   11.06    &  13.24      &    1.07    \\	
    NGC3198    & 10.30	   &	10.21	&  10.36	& 	  10.18	&     10.06		&  10.28	& 	 10.03     &    11.68       &    11.62  &   11.75	 &     11.55        & 	  11.50	 &   11.61    &   1.48      &    0.82    \\	
    NGC3521    & 10.64	   &	10.59	&  10.68	& 	  10.61	&     10.51		&  10.67	& 	  9.61     &    12.31       &    11.88  &   12.79	 &     11.69        & 	  11.45	 &   11.90    &   0.28      &    0.69    \\	
    NGC3726    & 10.44	   &	10.35	&  10.52	& 	  10.37	&     10.26		&  10.47	& 	  9.81     &    11.76       &    11.58  &   11.95	 &     11.49        & 	  11.38	 &   11.61    &   2.95      &    3.60    \\	
    NGC3741    &  7.16	   &	 6.97	&   7.34	& 	   7.24	&      7.05		&   7.43	& 	  8.26     &    10.46       &    10.26  &   10.70	 &     10.35        & 	  10.18	 &   10.52    &   1.58      &    1.04    \\	
    NGC3769    &  9.90	   &	 9.78	&  10.00	& 	   9.85	&      9.72		&   9.96	& 	  9.74     &    11.42       &    11.27  &   11.58	 &     11.26        & 	  11.14	 &   11.38    &   0.91      &    0.69    \\	
    NGC3893    & 10.47	   &	10.40	&  10.52	& 	  10.41	&     10.31		&  10.49	& 	  9.76     &    12.06       &    11.82  &   12.37	 &     11.62        & 	  11.49	 &   11.75    &   1.40      &    0.41    \\	
    NGC3917    & 10.21	   &	10.06	&  10.46	& 	  10.08	&      9.91		&  10.25	& 	  9.27     &    11.59       &    10.40  &   11.91	 &     11.31        & 	  11.13	 &   11.45    &   4.11      &    0.89    \\	
    NGC3972    &  9.84	   &	 9.68	&   9.99	& 	   9.87	&      9.71		&  10.02	& 	  9.08     &    11.95       &    11.56  &   12.37	 &     11.30        & 	  11.10	 &   11.48    &   1.43      &    0.76    \\	
    NGC3992    & 11.25	   &	11.19	&  11.30	& 	  11.05	&     10.93		&  11.15	& 	 10.22     &    12.16       &    12.05  &   12.29	 &     12.03        & 	  11.94	 &   12.16    &   0.85      &    0.61    \\	
    NGC4010    &  9.83	   &	 9.68	&   9.97	& 	   9.84	&      9.69		&   9.97	& 	  9.45     &    11.79       &    11.48  &   12.14	 &     11.33        & 	  11.16	 &   11.49    &   2.97      &    2.37    \\	
    NGC4013    & 10.60	   &	10.54	&  10.65	& 	  10.48	&     10.38		&  10.56	& 	  9.47     &    11.96       &    11.81  &   12.14	 &     11.70        & 	  11.60	 &   11.85    &   1.31      &    2.34    \\	
    NGC4085    &  9.77	   &	 9.65	&   9.88	& 	   9.79	&      9.68		&   9.88	& 	  9.13     &    12.30       &    11.82  &   12.84	 &     11.47        & 	  11.30	 &   11.66    &   5.05      &    2.39    \\	
    NGC4088    & 10.55	   &	10.47	&  10.62	& 	  10.51	&     10.41		&  10.59	& 	  9.91     &    11.73       &    11.50  &   11.98	 &     11.44        & 	  11.29	 &   11.59    &   0.66      &    0.97    \\	
    NGC4100    & 10.62	   &	10.55	&  10.68	& 	  10.54	&     10.43		&  10.62	& 	  9.49     &    11.73       &    11.54  &   11.94	 &     11.53        & 	  11.39	 &   11.69    &   1.31      &    0.93    \\	
    NGC4138    & 10.50	   &	10.43	&  10.56	& 	  10.45	&     10.36		&  10.52	& 	  9.17     &    11.52       &    11.19  &   11.84	 &     11.36        & 	  11.15	 &   11.57    &   1.71      &    1.76    \\	
    NGC4157    & 10.64	   &	10.56	&  10.71	& 	  10.57	&     10.47		&  10.66	& 	  9.91     &    11.96       &    11.76  &   12.18	 &     11.63        & 	  11.51	 &   11.75    &   0.56      &    0.93    \\	
    NGC4183    &  9.86	   &	 9.68	&  10.01	& 	   9.82	&      9.64		&   9.99	& 	  9.54     &    11.18       &    11.00  &   11.36	 &     11.06        & 	  10.91	 &   11.20    &   0.19      &    0.17    \\	
    NGC4214    &  8.95	   &	 8.79	&   9.08	& 	   8.90	&      8.69		&   9.08	& 	  8.68     &    11.31       &    10.92  &   11.81	 &     10.99        & 	  10.79	 &   11.17    &   0.91      &    0.48    \\	
    NGC4217    & 10.33	   &	10.26	&  10.40	& 	  10.29	&     10.20		&  10.36	& 	  9.40     &    12.41       &    12.13  &   12.77	 &     11.74        & 	  11.64	 &   11.86    &   5.27      &    2.52    \\	
    NGC4559    &  9.92	   &	 9.79	&  10.03	& 	   9.88	&      9.74		&  10.00	& 	  9.76     &    11.39       &    11.23  &   11.58	 &     11.20        & 	  11.07	 &   11.32    &   0.39      &    0.45    \\	
    NGC5033    & 10.74	   &	10.66	&  10.80	& 	  10.65	&     10.56		&  10.72	& 	 10.05     &    11.94       &    11.88  &   11.99	 &     11.82        & 	  11.78	 &   11.88    &   5.21      &    2.51    \\	
    NGC5055    & 10.70	   &	10.66	&  10.72	& 	  10.66	&     10.56		&  10.74	& 	 10.06     &    11.83       &    11.81  &   11.86	 &     11.81        & 	  11.77	 &   11.84    &   3.12      &    2.92    \\	
    NGC5371    & 11.22	   &	11.16	&  11.27	& 	  11.07	&     11.00		&  11.14	& 	 10.04     &    11.61       &    11.49  &   11.72	 &     11.99        & 	  11.84	 &   12.12    &   9.65      &    1.55    \\	
    NGC5585    &  8.92	   &	 8.80	&   9.03	& 	   9.22	&      9.16		&   9.27	& 	  9.22     &    11.42       &    11.27  &   11.61	 &     11.13        & 	  11.02	 &   11.24    &   6.25      &    5.22    \\	
    NGC5907    & 11.07	   &	11.00	&  11.11	& 	  10.83	&     10.75		&  10.90	& 	 10.32     &    12.02       &    11.94  &   12.12	 &     12.08        & 	  11.93	 &   12.23    &   6.39      &    3.13    \\	
    NGC5985    & 11.43	   &	11.32	&  11.49	& 	  11.04	&     10.86		&  11.21	& 	 10.06     &    11.99       &    11.72  &   12.18	 &     12.17        & 	  12.02	 &   12.32    &   8.09      &    2.44    \\	
    NGC6015    & 10.39	   &	10.33	&  10.43	& 	  10.21	&     10.12		&  10.29	& 	  9.76     &    11.67       &    11.54  &   11.83	 &     11.52        & 	  11.43	 &   11.61    &   8.45      &    7.44    \\	
    NGC6195    & 11.29	   &	11.26	&  11.31	& 	  11.12	&     10.99		&  11.24	& 	 10.32     &    12.13       &    11.94  &   12.34	 &     12.00        & 	  11.78	 &   12.19    &   3.47      &    3.43    \\	
    NGC6503    &  9.78	   &	 9.71	&   9.85	& 	   9.74	&      9.62		&   9.83	& 	  9.24     &    11.31       &    11.24  &   11.38	 &     11.25        & 	  11.17	 &   11.36    &   1.86      &    1.56    \\	
    NGC6674    & 11.30	   &	11.26	&  11.34	& 	  10.97	&     10.88		&  11.04	& 	 10.50     &    12.41       &    12.32  &   12.51	 &     12.45        & 	  12.37	 &   12.51    &   6.37      &    7.61    \\	
    NGC6946    & 10.48	   &	10.43	&  10.52	& 	  10.45	&     10.37		&  10.50	& 	  9.75     &    11.84       &    11.65  &   12.09	 &     11.49        & 	  11.37	 &   11.62    &   1.99      &    1.58    \\	
    NGC7331    & 10.97	   &	10.93	&  11.00	& 	  10.86	&     10.75		&  10.93	& 	 10.04     &    12.40       &    12.24  &   12.58	 &     12.01        & 	  11.92	 &   12.13    &   0.80      &    2.54    \\	
    NGC7814    & 10.68	   &	10.63	&  10.73	& 	  10.59	&     10.50		&  10.66	& 	  9.02     &    12.30       &    12.10  &   12.56	 &     11.87        & 	  11.77	 &   12.02    &   1.79      &    0.55    \\	
   UGC00128    &  9.82	   &	 9.63	&   9.98	& 	   9.71	&      9.54		&   9.86	& 	  9.87     &    11.56       &    11.53  &   11.59	 &     11.52        & 	  11.48	 &   11.57    &   3.28      &    3.85    \\	
   UGC00191    &  9.12	   &	 8.96	&   9.26	& 	   9.13	&      8.94		&   9.29	& 	  9.12     &    10.94       &    10.86  &   11.04	 &     10.74        & 	  10.63	 &   10.85    &   3.85      &    2.69    \\	
   UGC00731    &  8.29	   &	 8.09	&   8.49	& 	   8.30	&      8.10		&   8.50	& 	  9.25     &    10.77       &    10.64  &   10.92	 &     10.63        & 	  10.51	 &   10.77    &   0.38      &    0.11    \\	
   UGC01230    &  9.70	   &	 9.50	&   9.90	& 	   9.68	&      9.48		&   9.87	& 	  9.80     &    11.20       &    11.01  &   11.38	 &     11.09        & 	  10.94	 &   11.23    &   1.20      &    0.38    \\	
   UGC01281    &  8.31	   &	 8.11	&   8.52	& 	   8.31	&      8.11		&   8.49	& 	  8.46     &    10.46       &    10.10  &   10.85	 &     10.49        & 	  10.20	 &   10.79    &   2.84      &    0.29    \\	
   UGC02259    &  9.18	   &	 8.97	&   9.38	& 	   9.07	&      8.87		&   9.24	& 	  8.69     &    10.81       &    10.71  &   10.92	 &     10.97        & 	  10.82	 &   11.09    &   1.42      &    1.01    \\	
   UGC02487    & 11.71	   &	11.66	&  11.75	& 	  11.40	&     11.26		&  11.51	& 	 10.25     &    12.59       &    12.53  &   12.67	 &     12.44        & 	  12.40	 &   12.49    &   5.37      &    5.04    \\	
   UGC02885    & 11.41	   &	11.35	&  11.46	& 	  11.14	&     10.98		&  11.31	& 	 10.60     &    12.60       &    12.47  &   12.76	 &     12.42        & 	  12.26	 &   12.59    &   1.56      &    3.69    \\	
   UGC02916    & 10.71	   &	10.69	&  10.74	& 	  10.47	&     10.34		&  10.65	& 	 10.36     &    12.17       &    12.00  &   12.37	 &     12.02        & 	  11.85	 &   12.16    &  11.32      &    6.60    \\	
   UGC02953    & 11.19	   &	11.16	&  11.21	& 	  11.10	&     11.02		&  11.16	& 	  9.88     &    12.32       &    12.26  &   12.40	 &     12.17        & 	  12.10	 &   12.24    &   7.33      &    6.15    \\	
   UGC03205    & 10.93	   &	10.88	&  10.96	& 	  10.73	&     10.62		&  10.82	& 	  9.98     &    12.10       &    11.97  &   12.27	 &     11.91        & 	  11.79	 &   12.05    &   3.53      &    3.04    \\	
   UGC03546    & 10.68	   &	10.62	&  10.73	& 	  10.60	&     10.53		&  10.67	& 	  9.42     &    11.98       &    11.86  &   12.14	 &     11.74        & 	  11.65	 &   11.85    &   2.08      &    0.82    \\	
   UGC03580    &  9.49	   &	 9.41	&   9.55	& 	   9.54	&      9.45		&   9.61	& 	  9.64     &    11.58       &    11.48  &   11.69	 &     11.39        & 	  11.31	 &   11.47    &   3.89      &    4.85    \\	
   UGC04278    &  8.81	   &	 8.62	&   8.98	& 	   8.84	&      8.66		&   9.01	& 	  9.04     &    11.23       &    10.88  &   11.63	 &     10.97        & 	  10.73	 &   11.21    &   3.11      &    1.51    \\	
   UGC04325    &  9.38	   &	 9.14	&   9.59	& 	   9.29	&      9.05		&   9.54	& 	  8.83     &    10.86       &    10.56  &   11.09	 &     10.83        & 	  10.56	 &   11.06    &   3.33      &    3.76    \\	
   UGC04483    &  6.85	   &	 6.66	&   7.03	& 	   6.94	&      6.74		&   7.14	& 	  7.50     &     9.26       &     8.97  &    9.62	 &      9.46        & 	   9.18	 &    9.79    &   0.92      &    0.60    \\	
   UGC04499    &  8.86	   &	 8.69	&   9.01	& 	   8.89	&      8.71		&   9.04	& 	  9.04     &    10.85       &    10.67  &   11.06	 &     10.66        & 	  10.53	 &   10.81    &   1.48      &    0.34    \\	
   UGC05005    &  9.30	   &	 9.12	&   9.47	& 	   9.30	&      9.12		&   9.48	& 	  9.49     &    11.03       &    10.78  &   11.26	 &     10.87        & 	  10.68	 &   11.06    &   1.99      &    1.88    \\	
   UGC05253    & 10.96	   &	10.94	&  10.97	& 	  10.96	&     10.77		&  10.99	& 	 10.21     &    12.20       &    12.11  &   12.30	 &     12.00        & 	  11.91	 &   12.14    &   3.41      &    2.57    \\	
   UGC05414    &  8.65	   &	 8.48	&   8.81	& 	   8.73	&      8.57		&   8.87	& 	  8.75     &    10.92       &    10.63  &   11.26	 &     10.64        & 	  10.45	 &   10.84    &   3.27      &    0.40    \\	
   UGC05716    &  8.47	   &	 8.30	&   8.63	& 	   8.51	&      8.33		&   8.69	& 	  9.03     &    10.83       &    10.77  &   10.89	 &     10.75        & 	  10.67	 &   10.83    &   2.02      &    2.46    \\	
   UGC05721    &  8.68	   &	 8.50	&   8.83	& 	   8.60	&      8.39		&   8.79	& 	  8.75     &    11.01       &    10.78  &   11.33	 &     10.94        & 	  10.75	 &   11.11    &   2.20      &    1.72    \\	
   UGC05764    &  7.75	   &	 7.54	&   7.96	& 	   7.74	&      7.53		&   7.95	& 	  8.21     &    10.31       &    10.22  &   10.41	 &     10.16        & 	  10.03	 &   10.36    &   7.85      &    5.02    \\	
   UGC05829    &  8.52	   &	 8.32	&   8.72	& 	   8.53	&      8.33		&   8.73	& 	  9.01     &    10.40       &    10.13  &   10.71	 &     10.37        & 	  10.14	 &   10.60    &   1.10      &    0.33    \\	
   UGC05918    &  8.15	   &	 7.95	&   8.36	& 	   8.17	&      7.96		&   8.37	& 	  8.47     &    10.03       &     9.79  &   10.31	 &     10.04        & 	   9.83	 &   10.25    &   0.39      &    0.04    \\	
   UGC05986    &  9.48	   &	 9.35	&   9.59	& 	   9.49	&      9.34		&   9.61	& 	  9.42     &    11.92       &    11.67  &   12.26	 &     11.25        & 	  11.14	 &   11.39    &   6.06      &    0.87    \\	
   UGC06399    &  9.12	   &	 8.93	&   9.31	& 	   9.13	&      8.94		&   9.32	& 	  8.82     &    11.20       &    10.92  &   11.52	 &     10.93        & 	  10.75	 &   11.11    &   1.03      &    0.06    \\	
   UGC06446    &  8.83	   &	 8.63	&   9.02	& 	   8.82	&      8.61		&   9.02	& 	  9.14     &    10.98       &    10.77  &   11.22	 &     10.85        & 	  10.68	 &   11.01    &   0.38      &    0.22    \\	
   UGC06614    & 10.67	   &	10.57	&  10.76	& 	  10.60	&     10.47		&  10.70	& 	 10.34     &    12.19       &    12.01  &   12.38	 &     11.81        & 	  11.69	 &   11.92    &   0.59      &    2.86    \\	
   UGC06667    &  8.93	   &	 8.73	&   9.13	& 	   8.93	&      8.73		&   9.13	& 	  8.90     &    11.37       &    11.15  &   11.65	 &     10.97        & 	  10.84	 &   11.11    &   1.63      &    0.28    \\	
   UGC06786    & 10.68	   &	10.62	&  10.72	& 	  10.54	&     10.40		&  10.64	& 	  9.70     &    12.27       &    12.16  &   12.41	 &     12.01        & 	  11.93	 &   12.11    &   1.92      &    0.99    \\	
   UGC06787    & 10.72	   &	10.67	&  10.76	& 	  10.63	&     10.50		&  10.70	& 	  9.70     &    12.21       &    12.13  &   12.29	 &     12.10        & 	  12.00	 &   12.22    &  28.04      &   26.56    \\	
   UGC06818    &  8.74	   &	 8.58	&   8.89	& 	   8.73	&      8.58		&   8.87	& 	  9.03     &    10.86       &    10.55  &   11.20	 &     10.86        & 	  10.63	 &   11.10    &   6.06      &    3.47    \\	
   UGC06917    &  9.54	   &	 9.37	&   9.69	& 	   9.56	&      9.39		&   9.71	& 	  9.30     &    11.45       &    11.23  &   11.72	 &     11.12        & 	  10.98	 &   11.27    &   0.94      &    0.28    \\	
   UGC06923    &  9.10	   &	 8.94	&   9.25	& 	   9.12	&      8.97		&   9.25	& 	  8.90     &    11.11       &    10.77  &   11.52	 &     10.87        & 	  10.67	 &   11.09    &   1.47      &    0.94    \\	
   UGC06930    &  9.74	   &	 9.55	&   9.91	& 	   9.72	&      9.53		&   9.90	& 	  9.51     &    11.17       &    10.97  &   11.39	 &     11.02        & 	  10.86	 &   11.17    &   0.33      &    0.19    \\	
   UGC06973    & 10.04	   &	10.00	&  10.08	& 	  10.04	&      9.97		&  10.09	& 	  9.24     &    12.95       &    12.46  &   13.52	 &     11.78        & 	  11.63	 &   11.93    &   1.37      &    1.12    \\	
   UGC06983    &  9.54	   &	 9.36	&   9.71	& 	   9.51	&      9.32		&   9.68	& 	  9.47     &    11.34       &    11.15  &   11.56	 &     11.13        & 	  10.99	 &   11.27    &   0.72      &    0.50    \\	
   UGC07125    &  9.06	   &	 8.90	&   9.21	& 	   9.07	&      8.90		&   9.22	& 	  9.66     &    10.40       &    10.28  &   10.53	 &     10.37        & 	  10.26	 &   10.48    &   1.83      &    0.91    \\	
   UGC07151    &  9.21	   &	 9.06	&   9.33	& 	   9.27	&      9.12		&   9.38	& 	  8.79     &    10.77       &    10.48  &   11.08	 &     10.49        & 	  10.28	 &   10.70    &   2.76      &    1.08    \\	
   UGC07261    &  9.05	   &	 8.87	&   9.20	& 	   9.04	&      8.85		&   9.20	& 	  9.14     &    10.77       &    10.51  &   11.07	 &     10.68        & 	  10.48	 &   10.86    &   0.20      &    0.08    \\	
   UGC07399    &  8.98	   &	 8.78	&   9.16	& 	   8.90	&      8.69		&   9.11	& 	  8.87     &    11.49       &    11.24  &   11.84	 &     11.22        & 	  11.04	 &   11.39    &   1.75      &    1.05    \\	
   UGC07524    &  9.13	   &	 8.93	&   9.31	& 	   9.16	&      8.97		&   9.35	& 	  9.25     &    10.93       &    10.73  &   11.16	 &     10.74        & 	  10.58	 &   10.89    &   1.32      &    0.36    \\	
   UGC07559    &  7.83	   &	 7.61	&   8.17	& 	   7.77	&      7.59		&   7.97	& 	  8.22     &     9.13       &     6.75  &    9.56	 &      9.56        & 	   9.17	 &    9.92    &   1.99      &    0.30    \\	
   UGC07603    &  8.30	   &	 8.13	&   8.46	& 	   8.40	&      8.21		&   8.56	& 	  8.41     &    11.00       &    10.71  &   11.38	 &     10.68        & 	  10.47	 &   10.89    &   1.71      &    0.79    \\	
   UGC07608    &  8.19	   &	 7.99	&   8.38	& 	   8.20	&      8.00		&   8.39	& 	  8.72     &    10.94       &    10.59  &   11.36	 &     10.76        & 	  10.52	 &   11.02    &   1.15      &    0.06    \\	
   UGC07690    &  8.84	   &	 8.71	&   8.95	& 	   8.85	&      8.69		&   8.96	& 	  8.59     &    10.22       &     9.90  &   10.54	 &     10.26        & 	   9.98	 &   10.49    &   0.54      &    0.75    \\	
   UGC08286    &  9.07	   &	 8.87	&   9.23	& 	   8.99	&      8.80		&   9.18	& 	  8.80     &    10.93       &    10.80  &   11.09	 &     10.79        & 	  10.65	 &   10.95    &   2.39      &    1.77    \\	
   UGC08490    &  8.95	   &	 8.77	&   9.10	& 	   8.89	&      8.68		&   9.08	& 	  8.85     &    10.84       &    10.70  &   11.02	 &     10.84        & 	  10.68	 &   10.98    &   0.53      &    0.37    \\	
   UGC08550    &  8.29	   &	 8.11	&   8.45	& 	   8.34	&      8.13		&   8.51	& 	  8.45     &    10.50       &    10.34  &   10.70	 &     10.42        & 	  10.26	 &   10.59    &   0.74      &    0.58    \\	
   UGC08699    & 10.49	   &	10.46	&  10.52	& 	  10.45	&     10.35		&  10.51	& 	  9.57     &    12.00       &    11.81  &   12.24	 &     11.67        & 	  11.54	 &   11.83    &   1.27      &    1.12    \\	
   UGC09037    & 10.19	   &	10.08	&  10.30	& 	  10.17	&     10.05		&  10.28	& 	 10.28     &    11.88       &    11.69  &   12.10	 &     11.53        & 	  11.41	 &   11.64    &   2.44      &    1.92    \\	
   UGC09133    & 11.17	   &	11.14	&  11.19	& 	  10.96	&     10.89		&  11.04	& 	 10.52     &    12.22       &    12.19  &   12.26	 &     12.25        & 	  12.19	 &   12.32    &   8.92      &    7.85    \\	
   UGC10310    &  9.07	   &	 8.87	&   9.27	& 	   9.10	&      8.89		&   9.30	& 	  9.07     &    10.65       &    10.38  &   10.93	 &     10.54        & 	  10.33	 &   10.73    &   0.63      &    0.37    \\	
   UGC11455    & 11.22	   &	11.15	&  11.27	& 	  11.05	&     10.94		&  11.15	& 	 10.12     &    12.60       &    12.42  &   12.80	 &     12.13        & 	  12.05	 &   12.20    &   5.60      &    1.93    \\	
   UGC11820    &  8.69	   &	 8.51	&   8.85	& 	   8.77	&      8.57		&   8.96	& 	  9.29     &    11.11       &    11.02  &   11.23	 &     11.01        & 	  10.93	 &   11.09    &   2.15      &    5.31    \\	
   UGC11914    & 11.02	   &	10.99	&  11.04	& 	  11.00	&     10.85		&  11.04	& 	  8.94     &    13.02       &    12.50  &   13.59	 &     11.90        & 	  11.66	 &   12.21    &   2.54      &    2.68    \\	
   UGC12506    & 11.12	   &	10.95	&  11.27	& 	  10.97	&     10.76		&  11.17	& 	 10.55     &    12.16       &    11.91  &   12.37	 &     11.96        & 	  11.78	 &   12.11    &   0.97      &    0.21    \\	
   UGC12632    &  8.90	   &	 8.70	&   9.09	& 	   8.90	&      8.70		&   9.10	& 	  9.24     &    10.71       &    10.55  &   10.88	 &     10.61        & 	  10.48	 &   10.75    &   0.40      &    0.09    \\	
   UGC12732    &  8.96	   &	 8.77	&   9.14	& 	   8.97	&      8.78		&   9.15	& 	  9.56     &    11.09       &    10.96  &   11.25	 &     10.93        & 	  10.82	 &   11.05    &   0.36      &    0.67    \\	
    UGCA281    &  8.07	   &	 7.88	&   8.27	& 	   8.22	&      8.01		&   8.40	& 	  7.79     &     9.78       &     9.23  &   10.32	 &      9.83        & 	   9.36	 &   10.25    &   0.92      &    0.35    \\	
    UGCA442    &  7.87	   &	 7.69	&   8.06	& 	   7.90	&      7.71		&   8.09	& 	  8.42     &    10.86       &    10.72  &   11.03	 &     10.55        & 	  10.45	 &   10.66    &   3.27      &    1.43    \\	
    UGCA444    &  6.85	   &	 6.65	&   7.06	& 	   6.86	&      6.66		&   7.06	& 	  7.82     &     9.57       &     9.18  &   10.01	 &      9.78        & 	   9.41	 &   10.20    &   0.57      &    0.16    \\	
     ddo101    &  8.43	   &	 8.23	&   8.62	& 	   8.45	&      8.25		&   8.65	& 	  7.25     &    10.91       &    10.60  &   11.30	 &     10.65        & 	  10.43	 &   10.87    &   0.43      &    0.20    \\	
     ddo133    &  7.48	   &	 7.28	&   7.67	& 	   7.49	&      7.29		&   7.69	& 	  8.00     &    10.56       &    10.19  &   11.00	 &     10.40        & 	  10.14	 &   10.68    &   0.76      &    0.12    \\	
     ddo154    &  6.90	   &	 6.70	&   7.09	& 	   6.91	&      6.71		&   7.11	& 	  8.40     &    10.48       &    10.31  &   10.69	 &     10.30        & 	  10.16	 &   10.45    &   2.29      &    0.88    \\	
     ddo168    &  7.72	   &	 7.53	&   7.90	& 	   7.73	&      7.54		&   7.92	& 	  8.40     &    10.70       &    10.36  &   11.09	 &     10.68        & 	  10.41	 &   10.97    &   3.80      &    0.79    \\	
      ddo50    &  8.06	   &	 7.85	&   8.27	& 	   8.08	&      7.87		&   8.30	& 	  8.77     &     9.72       &     9.41  &   10.00	 &      9.79        & 	   9.49	 &   10.05    &   1.28      &    1.28    \\	
      ddo52    &  7.70	   &	 7.51	&   7.89	& 	   7.71	&      7.51		&   7.91	& 	  8.34     &    10.38       &    10.11  &   10.70	 &     10.34        & 	  10.12	 &   10.57    &   1.42      &    0.17    \\	
      ddo87    &  7.50	   &	 7.30	&   7.70	& 	   7.51	&      7.32		&   7.71	& 	  8.31     &    10.39       &    10.14  &   10.69	 &     10.31        & 	  10.11	 &   10.53    &   1.17      &    0.23    \\	
    ngc2366    &  7.83	   &	 7.63	&   8.03	& 	   7.83	&      7.63		&   8.03	& 	  8.80     &    10.52       &    10.27  &   10.82	 &     10.48        & 	  10.28	 &   10.71    &   4.33      &    1.15    \\	
        wlm    &  7.20	   &	 7.01	&   7.39	& 	   7.22	&      7.02		&   7.43	& 	  7.74     &    10.18       &     9.86  &   10.54	 &     10.08        & 	   9.82	 &   10.37    &   2.94      &    0.84    \\	

\end{longtable}




\begin{figure*}[h!]
\includegraphics[width=9.5cm]{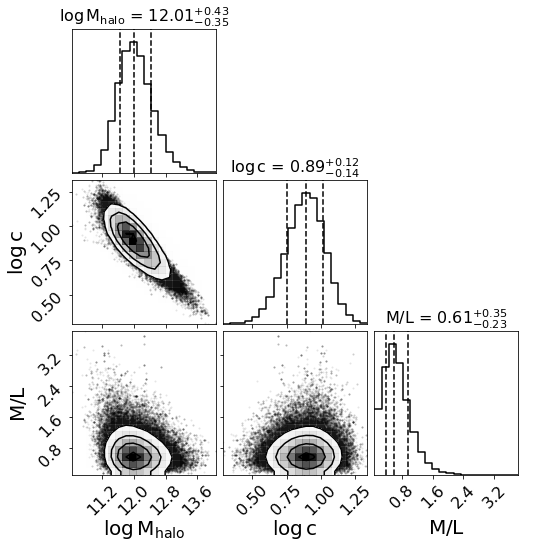}
\includegraphics[width=9.5cm]{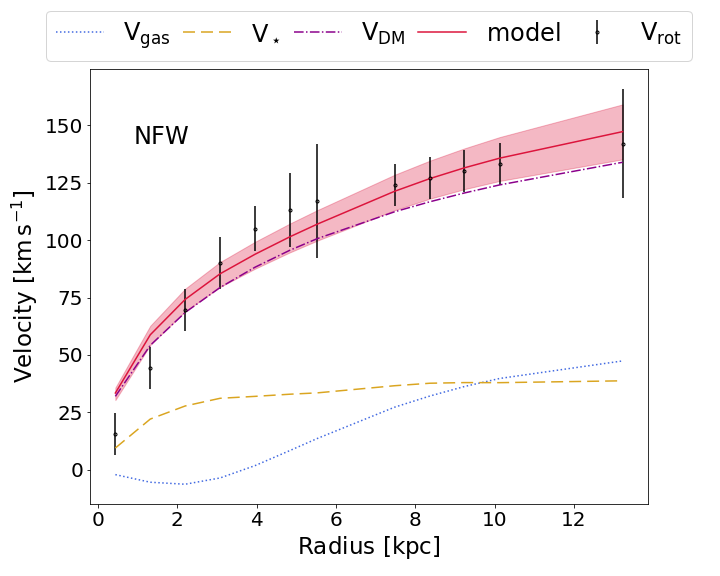}\\

\includegraphics[width=9.5cm]{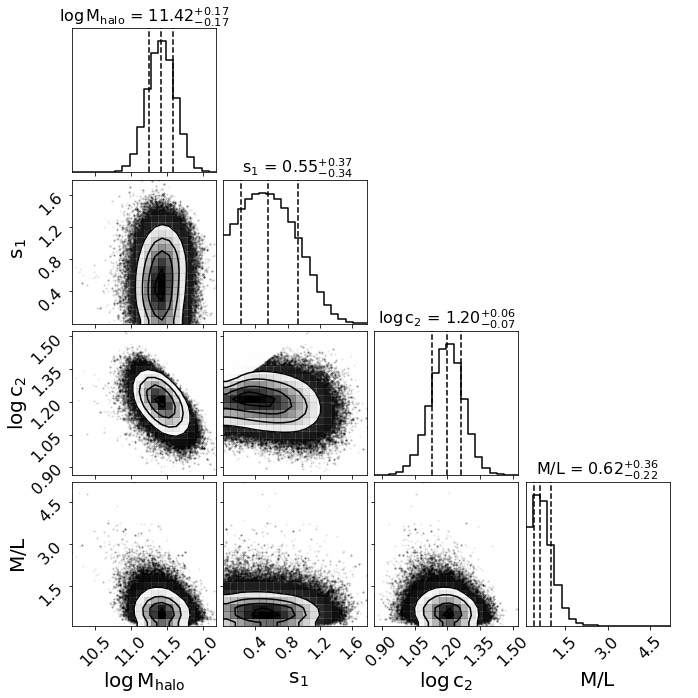}    
\includegraphics[width=9.5cm]{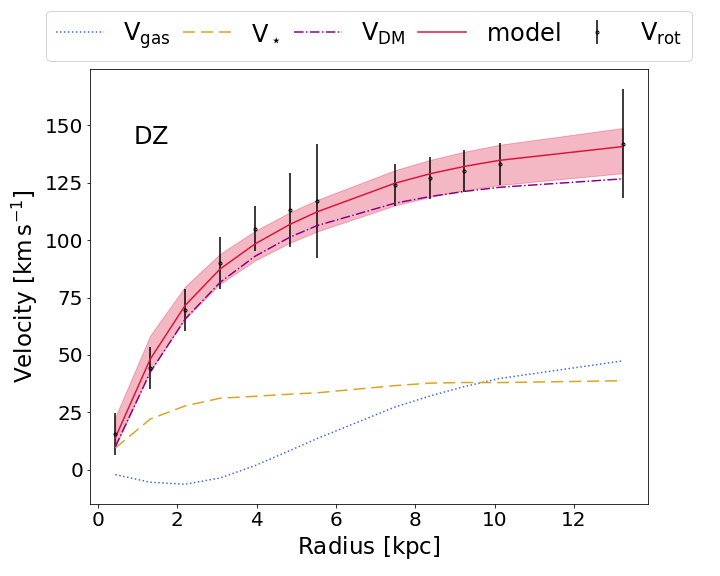}
\caption{Galaxy F568-1. Top row: NFW; left panel: Posterior distributions and mass model in right panel. Bottom row: DZ; left panel: Posterior distributions and mass model in right panel.}.
    \label{fig:NFW and DZ mass models}
\end{figure*}

\end{document}